\documentclass[reprint,aps,pra,superscriptaddress]{revtex4-1}

\usepackage{graphicx}
\usepackage{amsmath}
\usepackage{amssymb}
\usepackage{dcolumn}
\usepackage{bm}
\usepackage{setspace}
\usepackage{color}
\usepackage[capitalize,nameinlink]{cleveref}

\begin{document}

\title{\LARGE Flow topology during multiplexed particle manipulation using a Stokes Trap}

\author{Anish Shenoy}
\affiliation
{
	Department of Mechanical Science and Engineering \\ University of Illinois at Urbana-Champaign, Urbana, IL, 61801
}

\author{Dinesh Kumar}
\affiliation
{
	Department of Chemical and Biomolecular Engineering \\ University of Illinois at Urbana-Champaign, Urbana, IL, 61801
}
\affiliation
{
Beckman Institute for Advanced Science and Technology \\ University of Illinois at Urbana-Champaign, Urbana, IL, 61801
}

\author{Sascha Hilgenfeldt}
\affiliation
{
	Department of Mechanical Science and Engineering \\ University of Illinois at Urbana-Champaign, Urbana, IL, 61801
}

\author{Charles M. Schroeder}
\email[To whom correspondence must be addressed: ]{cms@illinois.edu}
\affiliation
{
	Department of Chemical and Biomolecular Engineering \\ University of Illinois at Urbana-Champaign, Urbana, IL, 61801
}
\affiliation
{
Beckman Institute for Advanced Science and Technology \\ University of Illinois at Urbana-Champaign, Urbana, IL, 61801
}
\affiliation
{
  Department of Materials Science and Engineering \\ University of Illinois at Urbana-Champaign, Urbana, IL, 61801
}

%\vspace{1 in}
\date{\today}

\begin{abstract}
Trapping and manipulation of small particles underlies many scientific and technological applications. Recently, the precise manipulation of multiple small particles was demonstrated using a Stokes trap that relies only on fluid flow without the need for optical or electric fields. Active flow control generates complex flow topologies around suspended particles during the trapping process, yet the relationship between the control algorithm and flow structure is not well understood. In this work, we characterize the flow topology during active control of particle trajectories using a Stokes trap. Our results show that optimal control of two particles unexpectedly relies on flow patterns with zero or one stagnation points, as opposed to positioning two particles using two distinct stagnation points. We characterize the sensitivity of the system with respect to the parameters in the control objective function, thereby providing a systematic understanding of the trapping process. Overall, these results will be useful in guiding applications involving the controlled manipulation of multiple colloidal particles and the precise deformation of soft particles in defined flow fields. 

\end{abstract}

\maketitle

In recent years, particle trapping methods have been used to study physical and biological phenomena with exquisite levels of precision. Particle trapping has been used for diverse applications including parallel manipulation of single particles \cite{chiou_massively_2005}, determination of DNA physical properties \cite{wang_stretching_1997}, and direct observation of viral DNA packaging by a molecular motor \cite{chemla_mechanism_2005}. Trapping methods rely on a variety of physical mechanisms and fundamental forces for manipulating particles, including optical fields \cite{ashkin_observation_1986,grier_revolution_2003}, magnetic fields \cite{gosse_magnetic_2002,sarkar_guide_2016}, electrical fields \cite{cohen_suppressing_2006,ropp_manipulating_2010}, acoustic forces \cite{hertz_standing-wave_1995,guo_three-dimensional_2016}, and fluidic forces \cite{petit_selective_2011,tanyeri_microfluidic-based_2011,tanyeri_manipulation_2013,shenoy_characterizing_2015,shenoy_stokes_2016,kumar_orientation_2019}. 

One of the most fundamental experiments in particle trapping focuses on measuring particle deformation in response to a controlled force \cite{gosse_magnetic_2002,wang_stretching_1997}. Flow-based trapping has long been used to study the deformation of suspended particles such as immiscible droplets. G.I. Taylor developed a four-roll mill for studying drop deformation in extensional flows \cite{taylor_formation_1934}, though this method required manual operation for trapping drops in flow. Several years later, Bentley and Leal \cite{bentley_computer-controlled_1986} developed a computer controlled version of the four-roll mill that enables the trapping of freely suspended drops for long times. Several researchers further developed microfluidic analogs of the four-roll mill to generate well-defined shear, extensional, and mixed flows \cite{hudson_microfluidic_2004,lee_microfluidic_2007}. The microfluidic four-roll mill was used to observe the deformation and dynamics of suspended particles, such as tank-treading, tumbling, and trembling dynamics of single vesicles in flow \cite{deschamps_dynamics_2009}. However, these microfluidic devices did not incorporate active flow control to confine particles over long time periods. 

In 2011, Schroeder and coworkers developed an automated hydrodynamic trap to confine single particles in free solution \cite{tanyeri_microfluidic-based_2011,tanyeri_manipulation_2013,shenoy_characterizing_2015}. This method relies on active flow control using an on-chip metering valve to modulate flow rates in a 4-channel cross-slot device. The initial version of the hydrodynamic trap offers a simple design with a linear control algorithm \cite{tanyeri_microfluidic-based_2011}, thereby producing a single stagnation point to confine individual particles in a planar extensional flow. Using this approach, the stagnation point is translated over micron-scale distances using an on-chip valve without affecting the principal axes of compression and extension, thereby generating a well-defined flow field that enables estimation of hydrodynamic forces on single suspended objects such as single polymer molecules \cite{schroeder2018single}. In 2016, the Stokes trap was developed as a general method to confine and manipulate multiple particles using the sole action of fluid flow \cite{shenoy_stokes_2016}. This technique utilizes a model predictive control formulation to modulate flow rates in an $N$-channel intersecting cross-slot device to achieve simultaneous center-of-mass trapping and precise manipulation of multiple particles. The Stokes trap was initially demonstrated with an $N$=6 channel device to manipulate two distinct micron-scale particles along pre-determined paths and for the directed assembly of two sticky particles using scheduled fluid flows \cite{shenoy_stokes_2016}. 

Although the Stokes trap enables manipulation of multiple particles, the flow structure and evolution of flow topology surrounding the suspended particles during the trapping process is not well understood. In theory, a six-channel microdevice is capable of generating two stagnation points (for trapping two distinct particles), though it is not known if the flow topology during particle trapping admits two stagnation points. Moreover, the evolution and location of putative stagnation points and streamlines during fluidic trapping experiments is generally not known a priori. The relation between the control algorithm used to determine optimal manipulation pathways and particle dynamics is not obvious, and the resulting particle trajectories may be complex depending on the trapping objective. From this perspective, achieving a clear understanding of the underlying flow topology would be useful for characterizing optimal manipulation techniques. For example, it is known that flow structures can interact with and influence the topology of molecular orientation fields in nematic systems \cite{giomi_cross-talk_2017}. Moreover, adhesive interactions between freely suspended vesicles in flow are known to depend on the stress profile induced on bilayer membrane by external flows. Consider the case of two vesicles that are on a trajectory for close approach or collision. Approaching vesicles experience compressional stresses, resulting in wrinkling of membrane surfaces, which reduces the effective contact area of adhesion, delays the film drainage process, and increases the time required for adhesion \cite{kantsler_vesicle_2007,turitsyn_wrinkling_2008,narsimhan_pearling_2015}. From this view, achieving a systematic understanding of how the underlying flow topologies evolve during particle manipulation and interaction experiments will be essential for quantitative descriptions of such processes. 

In this work, we systematically study flow topologies during particle manipulation experiments using a Stokes trap. In particular, we characterize flow topology during three canonical manipulation scenarios for two particles, including the movement of two particles towards each other, away from each other, and the interchange of particle positions. For each case, particle trajectories are tracked using active flow control experiments and directly compared to simulation results. Our results show that the experimentally determined flow rates are in good agreement with simulation results for several different manipulation scenarios. In addition, we systematically vary the control parameters in the objective function to understand the influence of the control algorithm on the time required for the particles to reach their target positions. Together, these studies elucidate the flow topology during particle manipulation and reveal the impact of the tunable control parameters on the trapping performance. Overall, these findings will help guide multiplexed flow-based trapping experiments involving particle collision, adhesion, and assembly in a systematic fashion.

\section{Controller formulation}
The flow model and the model predictive control (MPC) scheme for multiplexed particle manipulation have been previously described \cite{shenoy_stokes_2016,kumar_orientation_2019}. The following section summarizes the salient features of these models for understanding flow topology during trapping.

\subsection{Flow model for the particle center-of-mass.}
\begin{figure}
\begin{center}
\includegraphics{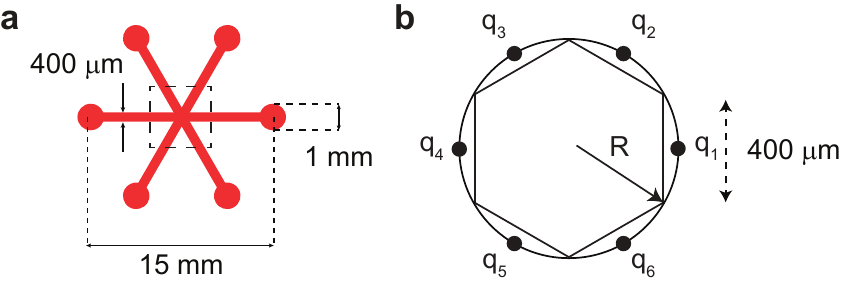}
\caption{\label{fig:device_design} Schematic of a six-channel microfluidic cross-slot device. Inlet/outet channels are approximated as point sources or sinks. (a) A six-channel device with the center portion forming a hexagonal cross-slot. (b) Position of the six point sources/sinks corresponding to each channel.}
\end{center}
\end{figure}

Consider the six-channel cross-slot device shown in \cref{fig:device_design}a. The intersection of the six channels creates a hexagonal cross-slot as shown in \cref{fig:device_design}b, with the vertices of the hexagon lying on a circle of radius $R$ which is equal to the width $W$ of a channel. The center of the circle is chosen to be the origin. Inlet/outlet channels in the cross-slot are assumed to be point sources/sinks \cite{schneider_algorithm_2011}, where $\bm{R}^i$ is the location of the i$^{th}$ source/sink, as shown schematically in \cref{fig:device_design}b.
\begin{align}
\bm{R}^i &=  W \left[\cos{\left((i-1)\frac{\pi}{3}\right)},\sin{\left((i-1)\frac{\pi}{3}\right)}\right]^T
\end{align}

Under low Reynolds number flow conditions, the height averaged 2-D velocity $\bm{u} \in \mathbb{R}^2$ within the cross-slot can be expressed as a linear superposition of the velocity fields caused by point sources using a 2-D Hele-Shaw approximation \cite{leal_advanced_2007}: 
\begin{align}
\bm{u} = \frac{1}{\pi  H} \sum^6_{i=1} \frac{(\bm{x}-\bm{R}^i)q_i}{\|\bm{x}-\bm{R}^i\|^2}
\label{eq:pointsource}
\end{align}
where $H$ is the height of the device, $\bm{x}\in \mathbb{R}^2$ is a position vector, $\bm{R}^i \in \mathbb{R}^2$ is the position vector corresponding to the $i^{th}$ point source/sink, and $\bm{q}\in\mathbb{R}^6$ is a vector whose $i^{th}$ element $q_i$ represents the volumetric flow rate through the $i^{th}$ point source. Prior work used computational fluid dynamics (CFD) simulations to demonstrate that proper selection of the dimensions of the microfluidic device ($H$ and $W$) can reduce the relative error between the velocity predicted by the model in \cref{eq:pointsource} and the CFD velocity to as low as 2\% \cite{shenoy_stokes_2016}. The microfluidic cross-slot is a closed system, and hence the flow rates $\bm{q}$ must satisfy mass conservation:
\begin{align}
\sum^6_{i=1}q_i = 0
\label{eq:sumflowrates}
\end{align}
Flow rates for a point source are taken to be positive when the fluid in the corresponding channel flows into the cross-slot and negative when the fluid in the channel leaves the cross-slot. Together, \cref{eq:pointsource} and \cref{eq:sumflowrates} determine the relationship between the imposed flow rates in the channels and the velocity field within the cross-slot. In the absence of external forces, a particle's 2-D velocity ($\dot{\bm{x}} \in \mathbb{R}^2$) is identical to the fluid velocity at the particle center-of-mass (COM) ($\bm{x} \in \mathbb{R}^2$):
\begin{align}
\dot{\bm{x}} = \bm{f}(\bm{x},\bm{q},\bm{R})\triangleq\frac{1}{\pi  H} \sum^6_{i=1} \frac{(\bm{x}-\bm{R}^i)q_i}{\|\bm{x}-\bm{R}^i\|^2}
\label{eq:singleparticlevelocity} 
\end{align}
where the $\triangleq$ symbol denotes a definition of the function.

\subsection{Model predictive control for particle manipulation.}
Using the linear velocity model for particle motion in \cref{eq:singleparticlevelocity}, we specify a control scheme for manipulating two particles from an initial position $\bm{X}_0 = [\bm{x}^T_{1,0}, \bm{x}^T_{2,0}]^T$ to a final position $\bm{X}_F = [\bm{x}^T_{1,F}, \bm{x}^T_{2,F}]^T$, where $\bm{x}_{i,0}$ indicates the $i^{th}$ particle's initial position and $\bm{x}_{i,F}$ indicates the target final position. A model predictive control (MPC) scheme is used to achieve precise manipulation of multiple particles \cite{mayne_constrained_2000,shenoy_stokes_2016,kumar_orientation_2019}. Using MPC, the flow rates required for particle manipulation are obtained by minimizing an objective function that determines an optimal balance between: (1) trajectories that move particles along the shortest path between the initial and final positions and (2) trajectories that require minimal flow rates but may require complicated paths. For the current sampling instant at time $t_0$, particle positions $\bm{X}_0$ are experimentally determined. Next, a future time horizon $T$ is defined and divided into $K$ equal intervals, such that $t_k = t_0+k\Delta,\ k=1\ldots K$, and $\bm{q}_k$ indicates the flow rates to be applied during $k^{th}$ interval $[t_{k-1}, t_{k})$. Given a series of flow rates throughout the time horizon $[\tilde{\bm{q}}_1,\tilde{\bm{q}}_2,\ldots,\tilde{\bm{q}}_K]$, the locations of all particles can be predicted subject to the imposed flow rates at the future $K$ sampling points. Future particle positions are denoted as $\tilde{\bm{X}}_k = [\tilde{\bm{x}}^T_{1,k},\tilde{\bm{x}}^T_{2,k}]^T$. 

\begin{figure}
\begin{center}
\includegraphics{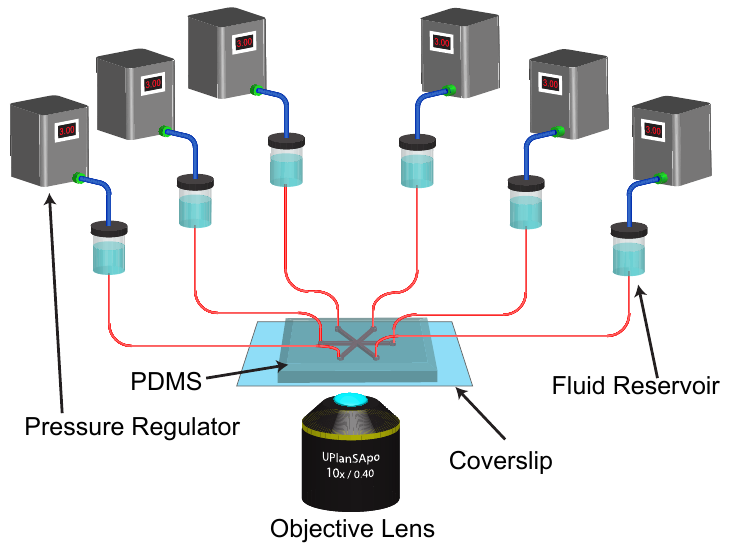}
\caption{\label{fig:controls_expsetup} Schematic of the experimental setup for the Stokes trap. Inlet/outlet channels of the microfluidic device are connected to fluidic reservoirs that are pressurized by regulators controlled by a custom LabVIEW program.}
\end{center}
\end{figure}

\begin{figure*}
\begin{center}
\includegraphics{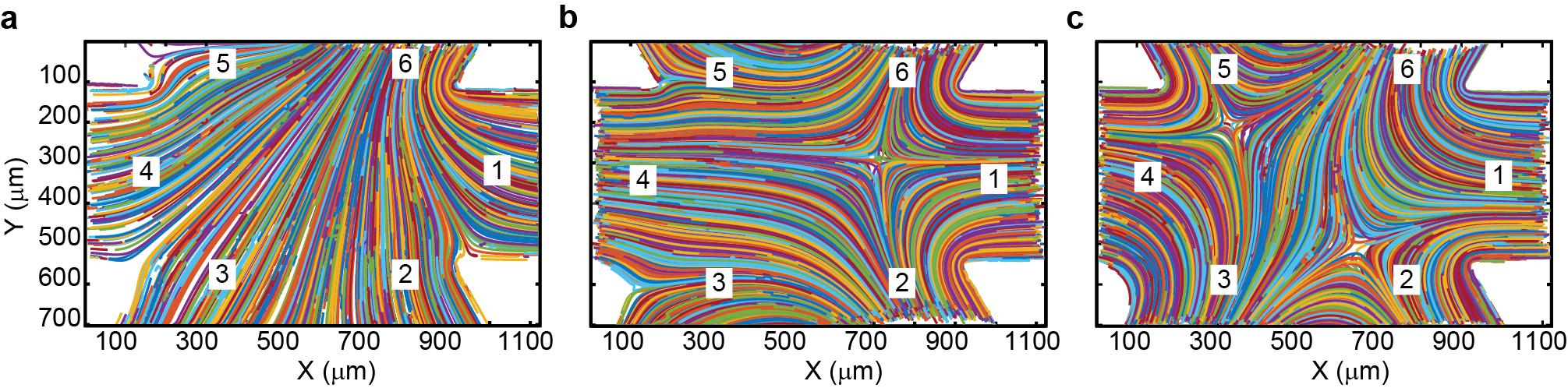}
\caption{\label{fig:three_topologies} Experimental flow topologies determined using particle tracking velocimetry (PTV). The numbers denote the channel number. Three primary flow topologies were determined in a six-channel cross-slot device exhibiting (a) zero, (b) one, and (c) two stagnation points.}
\end{center}
\end{figure*}

Using the MPC formulation, it is desired to obtain a set of flow rates that simultaneously minimize the distance travelled by each particle and the flow rates during the trapping event. To accomplish this task, an objective function $\mathcal{J}$ is minimized with respect to the particle trajectories $\{\tilde{\bm{X}}_k\}^{K}_{k=1}$ and flow rates $\{\tilde{\bm{q}}_k\}^{K}_{k=1}$:
\begin{subequations}
\begin{align}
\mathcal{J} &=\sum^{K-1}_{k=0} \left\{\|\tilde{\bm{X}}_k-\bm{X}_F\|^2 \nonumber +  \beta \|\tilde{\bm{q}}_k\|^2 \right\} \nonumber \\
& + \gamma \left( \|\tilde{\bm{X}}_K-\bm{X}_F\|^2 \right)\\
\text{s.t.} \quad & \frac{d \tilde{\bm{X}}}{dt}= [\bm{f}^T(\tilde{\bm{x}}_1,\tilde{\bm{q}}_k,\bm{R}),\bm{f}^T(\tilde{\bm{x}}_2,\tilde{\bm{q}}_k,\bm{R})]^T, \tilde{\bm{X}}_k = \bm{X}_0 \label{eq:mpcsysmodel}\\
&\sum^6_{i=1}   \tilde{q}_{k,i} =0 \quad \forall \ k = 1,\ldots, K
\end{align}
\label{eq:singleparticlecontrol}
\end{subequations}
where $\beta$ and $\gamma$ are control parameters or weights that penalize high flow rates and large deviations at the end of the time horizon. In general, $\beta$ and $\gamma$ are tuned by the user to achieve the desired speed of response during trapping experiments. The objective function in \cref{eq:singleparticlecontrol} is minimized at each sampling instant to obtain the set of flow rates for the horizon $[\tilde{\bm{q}}_1,\tilde{\bm{q}}_2,\ldots,\tilde{\bm{q}}_K]$, but only the flow rates corresponding to the first interval $\tilde{\bm{q}}_1$ are applied to the system. In the next sampling instant, the particle positions are resampled again and the process is repeated. The objective function defined by \cref{eq:singleparticlecontrol} relies on a non-linear model, which complicates the determination of a closed-form solution. For experimental implementation of the MPC scheme, we use the toolkit for Automatic Control and Dynamic Optimization (ACADO) \cite{houska_acado_2011,quirynen_autogenerating_2015}, which efficiently minimizes the objective function using numerical optimization methods and generates high performance C++ code to enable millisecond time solutions of \cref{eq:singleparticlecontrol}. Together, this approach enables real-time implementation of the non-linear model in optical microscopy and imaging experiments. 

\section{Methods}
\subsection{Experimental setup.}
Standard techniques in soft-lithography are used to fabricate the microfluidic device shown in \cref{fig:device_design} (Supplementary Materials) \cite{qin_soft_2010}. Microfluidic devices are mounted on the stage of an inverted microscope as shown in \cref{fig:controls_expsetup}. The flow setup consists of 6 pressure regulators (MPV Series, Proportion Air) connected to a nitrogen source with an output pressure of 30 psi. The pressure regulators accept an analog input voltage ranging from 0-10 V, provide a linearly dependent output of 0-5 psi, and can measure the output pressure through an analog output. The regulators are actuated through a computer-controlled LabVIEW data acquisition device with analog input and output capabilities (NI 9264, NI9205 and NI9174, National Instruments). Each output pressure line from the regulators is connected to a fluidic reservoir containing the sample (Elveflow XS fluid reservoir), which is connected to the microfluidic device using FEP and PEEK tubing (IDEX Health and Science). Tube dimensions (inner diameter and length) are chosen such the fluidic resistance within the tubing is significantly larger than the fluidic resistance of the microfluidic channels. The microfluidic device is then mounted on an inverted microscope (Olympus IX71) equipped with a mercury lamp for fluorescent imaging, a 10x magnification objective lens, and a charge-coupled device (CCD) camera (PointGrey GS3 23S6M). Unless otherwise noted, all experiments, were performed using fluorescent polystyrene beads (2.2 $\mu$m, Nile Red, Spherotech) suspended in an aqueous glycerol solution with a viscosity of 0.140 Pa-s measured at 25 $^o$C. The microfluidic device is situated such that the center of the cross-slot is defined as the origin in the laboratory reference frame.
\begin{figure}
\begin{center}
\includegraphics{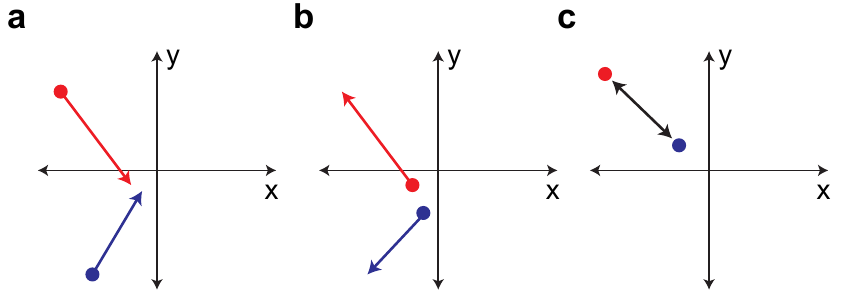}
\caption{\label{fig:manipulation_experiments} Schematic representation of particle manipulation experiments for studying flow topology. Red circles represent particle 1 and blue circles represents particle 2. Particles are trapped at the locations of the circle. Manipulation experiments are performed by (a) moving set points closer to each other, (b) moving set points away from each other, and (c) switching set point positions.}
\end{center}
\end{figure}

\begin{figure*}
\begin{center}
\includegraphics{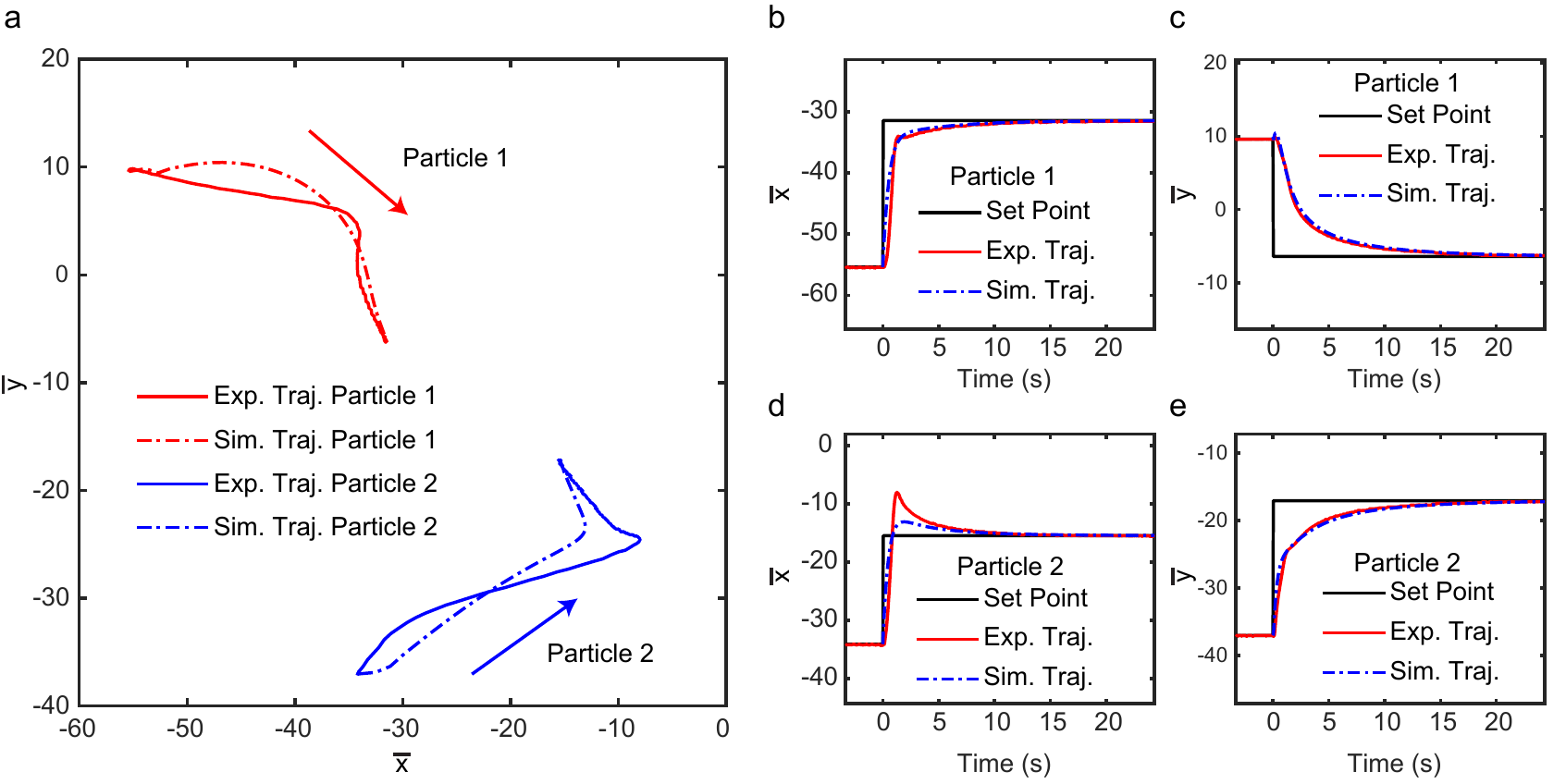}
\caption{\label{fig:converge_trajectories} Particle trajectories for the case of two particles moving towards each other (corresponding to the schematic in \cref{fig:manipulation_experiments}a). Two particles are first trapped and then at time $t$=0, the trapping point for both particles is changed to move them closer to each other. (a) Particle trajectories from experiments and simulations where the arrows are drawn to show the overall direction of movement of each particle. (b) $x$-trajectory of particle 1 from experiments and simulations. (b) $y$-trajectory of particle 1 from experiments and simulations. (d) $x$-trajectory of particle 2 from experiments and simulations. (e) $y$-trajectory of particle 2 from experiments and simulations.}
\end{center}
\end{figure*}

\subsection{Controller implementation.} 
Particle manipulation experiments are performed using a custom code written in MATLAB (Mathworks) and LabVIEW (National Instruments) \cite{shenoy_stokes_2016,kumar_orientation_2019}. Briefly, at each sampling instant, LabVIEW acquires an image (single frame) from the CCD camera and performs binary thresholding, followed by particle localization to determine the centroid positions of particles within a region of interest defined by the user. Within a region of interest, the particles nearest to the two set points are selected, and the center-of-mass (COM) coordinates of these particles are transformed from the camera reference frame to the laboratory reference frame and scaled to dimensionless quantities \cite{kumar_orientation_2019}. Scaling is used to avoid numerical underflow; in brief, particle coordinates are on the order of few hundred microns and flow rates are on the order of a few $\mu L /hr$, and determining the minima for the objective function in \cref{eq:singleparticlecontrol} can be strongly influenced by numerical error in the optimization algorithm. In this way, non-dimensionalization scales all quantities to be approximately $\mathcal{O}(1)$. All lengths are scaled by the particle diameter $d = 2.2 \ \mu$m, the time is scaled by a characteristic inverse strain rate $\dot{\epsilon}^{-1}=t_s=1$. Thus, \cref{eq:singleparticlecontrol} is solved using the dimensionless particle coordinates to obtain the dimensionless flow rates to be imposed on each channel. 

In order to calculate the corresponding pressures for each desired flow rate, the fluidic setup is converted to a fluidic circuit, assigning a reference pressure $P_0$ to the center of the cross-slot and a fluidic resistance $Z_i$ to the fluidic path from the $i^{th}$ reservoir to the cross-slot. The pressure $P_i$ to be applied to the $i^{th}$ fluid reservoir is calculated using the following ohmic relation between the pressure and the flow rate $q_i$:
\begin{align}
P_i = P_0 + q_i Z_i
\label{eq:pressure_to_flow}
\end{align}

\begin{figure}
\begin{center}
\includegraphics{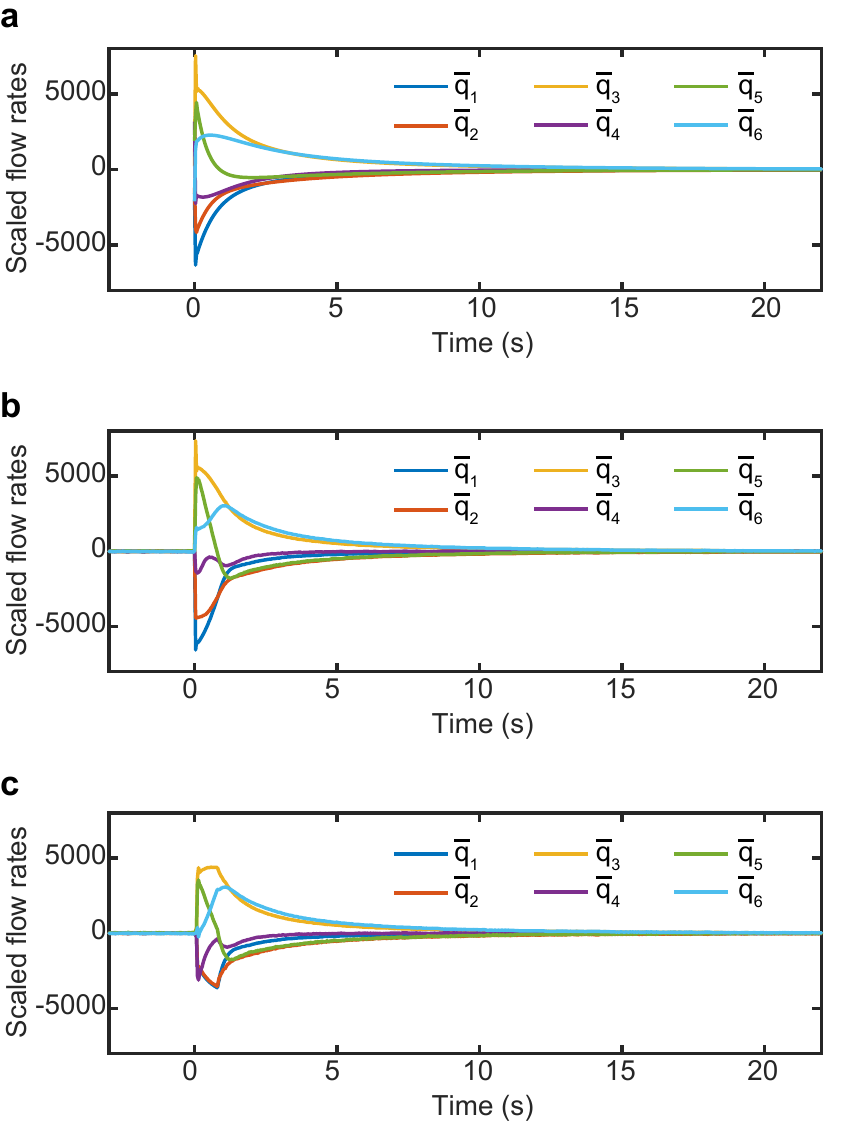}
\caption{\label{fig:converge_flowrates} Flow rates for the case of two particles moving towards each other, corresponding to the experiment and simulation shown in \cref{fig:converge_trajectories}. (a) Flow rates from simulations when carried out with identical values of the initial and final particle positions and control weights as in experiments. The simulation begins at time $t$=0. (b) Applied flow rates calculated by the control algorithm during the experiment, which are converted to pressure values and used to control the pressure regulators. (c) Actual flow rates back-calculated from the measured pressures determined by the pressure regulators in experiments.}
\end{center}
\end{figure}

Identical tubing and fittings are used for each channel, and the fluidic resistance due to the tubing is much larger than the microfluidic channels. We therefore assume the same value of the fluidic resistance $Z_i$ for the fluidic path from each reservoir to the cross-slot. Based on the viscosity of the fluid and the dimensions of the tubing and channels, $Z_i$ was found to be $1.4 \times 10^{15}$ Pa-s/m${^{3}}$. $P_0$ is generally set to be in the range of 2.5-3 psi, which is chosen to be in the intermediate range of the maximum of 5 psi for each pressure regulator, thereby maximizing the range of pressure variations in the positive and negative direction for controlling flow rates. Finally, actual pressure values are calculated using the process described above and converted to the appropriate analog voltage values, which is then used to drive fluid flow within the cross-slot device. The entire process repeats with a feedback time loop constant of 33 msec.

\section{Results And Discussion}

\subsection{Flow topologies in a six-channel device.}
We began by experimentally characterizing permissible flow topologies in a six-channel device. An aqueous suspension of fluorescently labeled microspheres are introduced into the cross-slot channel, and the pressures are varied in the six-channel device to perform particle tracking experiments. In \cref{fig:three_topologies}, the three permissible flow topologies are depicted for an $N$=6 channel device based on particle tracking velocimetry (PTV) experiments. The three flow topologies are characterized based on the number of stagnation points in the resulting flow pattern. In the first flow topology (\cref{fig:three_topologies}a), five inlet streams and 1 outlet stream give rise to zero stagnation points within the cross-slot. \cref{fig:three_topologies}b,c show two situations in which the pattern of inflow and outflow channels are the same. In the second flow topology (\cref{fig:three_topologies}b), in clockwise direction beginning with the horizontal channel on the right, channels 1, 3, 4, and 5 are inlet channels while channels 2 and 6 are outlet channels. In this configuration, there is a single stagnation point created in the cross-slot. In the third flow topology (\cref{fig:three_topologies}c), there are two stagnation points created with channels 1, 3, 4, and 5 being inlet channels, and channels 2, and 6 being outlet channels. Both \cref{fig:three_topologies}b,c have the same inlet and outlet channels, but by changing the quantitative flow rates in each channel, flow topologies with one or two stagnation points are possible.

\begin{figure}
\begin{center}
\includegraphics{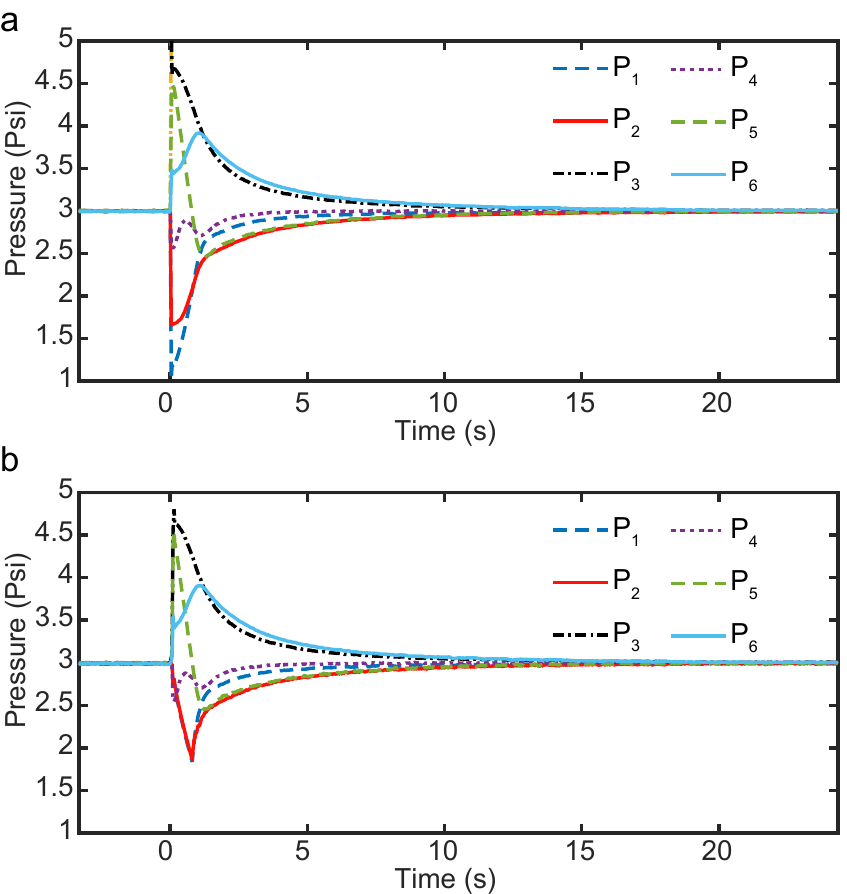}
\caption{\label{fig:converge_pressures} Experimentally applied and measured pressures for each of the 6 channels for the case of two particles moving towards each other, corresponding to the experiment and simulation shown in \cref{fig:converge_trajectories}. (a) Applied pressures applied during the experiment. (b) Actual pressures measured by the regulator during the experiment.}
\end{center}
\end{figure}

\subsection{Flow topologies during particle manipulation.}
The Stokes trap allows for different types of trapping or manipulation experiments. In one type of experiment, two particles can be trapped at pre-determined separate locations inside the cross-slot. In this way, two particles are essentially confined at the location of two distinct stagnation points. In a second type of experiment, two particles can be independently moved along pre-determined trajectories from initial to final positions, where particle trajectories are selected in an optimal manner as characterized by \cref{eq:singleparticlecontrol}. At the end of the particle translation experiment, the flow field will typically vanish, however, trapping can be re-initiated to the new particle positions if desired. 

\begin{figure}
\begin{center}
\includegraphics{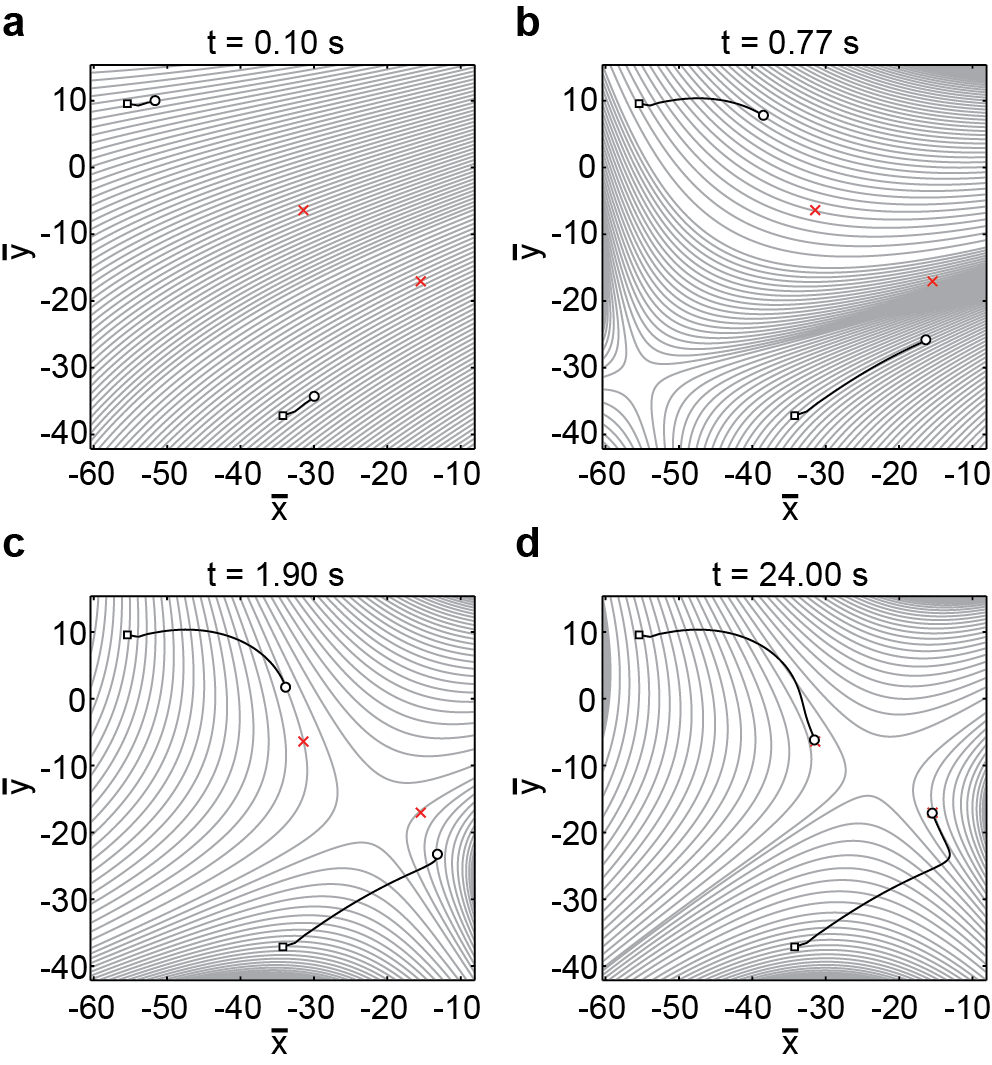}
\caption{\label{fig:converge_streamlines} Streamlines at different instants in time during the manipulation process as particles are brought closer to each other, corresponding to the trapping process shown in \cref{fig:converge_trajectories}. Streamlines are shown at (a) t = 0.10 s, (b) t = 0.77 s, (c) t = 1.90 s, and (d) t = 24.00 s.}
\end{center}
\end{figure}

In the following, we examine flow topology during the multiple particle manipulation experiment, wherein the initial and final positions for particle trajectories are intentionally chosen such that the emergent flow topologies are not a result of the symmetry of the device. In this way, we perform three different particle manipulation protocols: (1) moving two particles closer to each other, (2) moving two particles away from each other, and (3) interchanging the positions of the particles, shown schematically in \cref{fig:manipulation_experiments}. 

\begin{figure*}
\begin{center}
\includegraphics{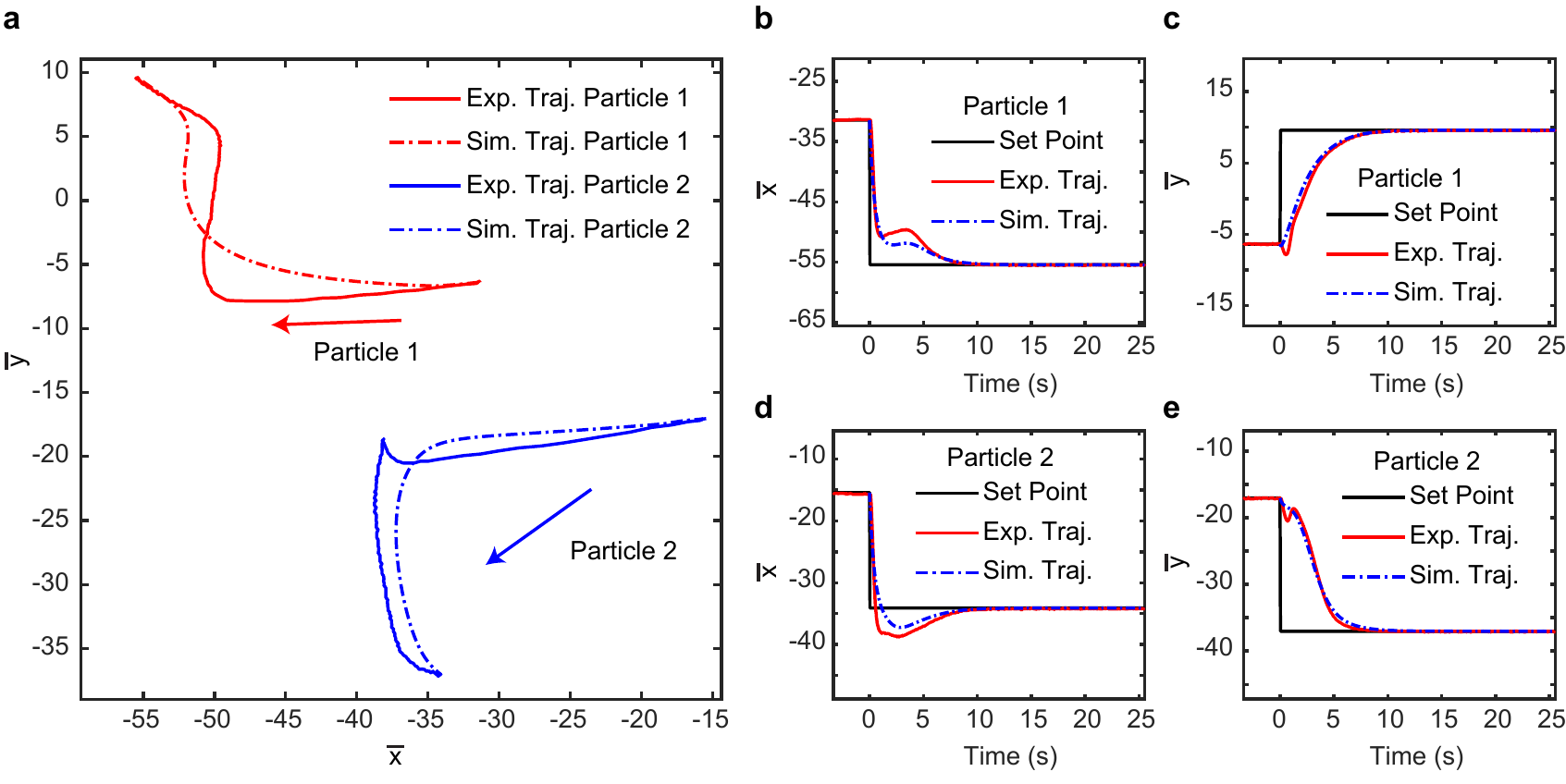}
\caption{\label{fig:diverge_trajectories} Particle trajectories for the case of two particles moving away from each other (corresponding to the schematic in \cref{fig:manipulation_experiments}b).Two particles are first trapped and then at time $t$=0, the trapping point for both particles is changed to move them away from each other. (a) Particle trajectories from experiments and simulations where the arrows are drawn to show the overall direction of movement of each particle. (b) $x$-trajectory of particle 1 from experiments and simulations. (b) $y$-trajectory of particle 1 from experiments and simulations. (d) $x$-trajectory of particle 2 from experiments and simulations. (e) $y$-trajectory of particle 2 from experiments and simulations.}
\end{center}
\end{figure*}

Flow topologies and particle trajectories are characterized using experiments and numerical solution of the flow model. For experiments, the particle position, applied flow rates, applied pressures, and measured pressures for each regulator are recorded at each sampling instance (sampled at a rate of 30 Hz). The control weights are set to $\beta=0.0001$ and $\gamma=1000$. For simulations, the initial and target positions of the two particles and the weight parameters in the control model are set identical to experimental values. In this way, the MPC algorithm \cref{eq:singleparticlecontrol} was used to determine the simulation flow rates for the next sampling interval, and \cref{eq:singleparticlevelocity} was numerically integrated using a suitable scheme (e.g. Runge-Kutta 45) to estimate particle position in the next sampling instant given these flow rates. This process was repeated until the the total simulation time matched the experimental duration. Brownian motion was not included in the model because particle Peclet numbers were generally larger than unity for all experiments. 

We began by studying particle trajectories for the case of two particles approaching each other during active control, which corresponds to the schematic in \cref{fig:manipulation_experiments}a.  \cref{fig:converge_trajectories}a shows particle trajectories from experiments and numerical simulations. Particle 1 was initially trapped at a coordinate of [-55, 10] and particle 2 was initially trapped at the coordinate [-34, -37], where positions are given in dimensionless units. At time $t$=0, particle 1's target location was set to [-31, -6] and particle 2's target location was set to [-16, -17]. The dashed lines indicate the simulation trajectory and the solid lines show the experimental trajectories. As shown in \cref{fig:converge_trajectories}a, good agreement is generally observed between particle trajectories from experiments and simulations. The individual `x' and `y' components corresponding to experimental and simulation trajectories for each particle are plotted in \cref{fig:converge_trajectories}b-e. In general, our results show that both components of the simulation trajectories agree well with the experimental trajectories, with both converging exponentially towards their target positions (Supplementary Material Video 1).

\begin{figure}
\begin{center}
\includegraphics{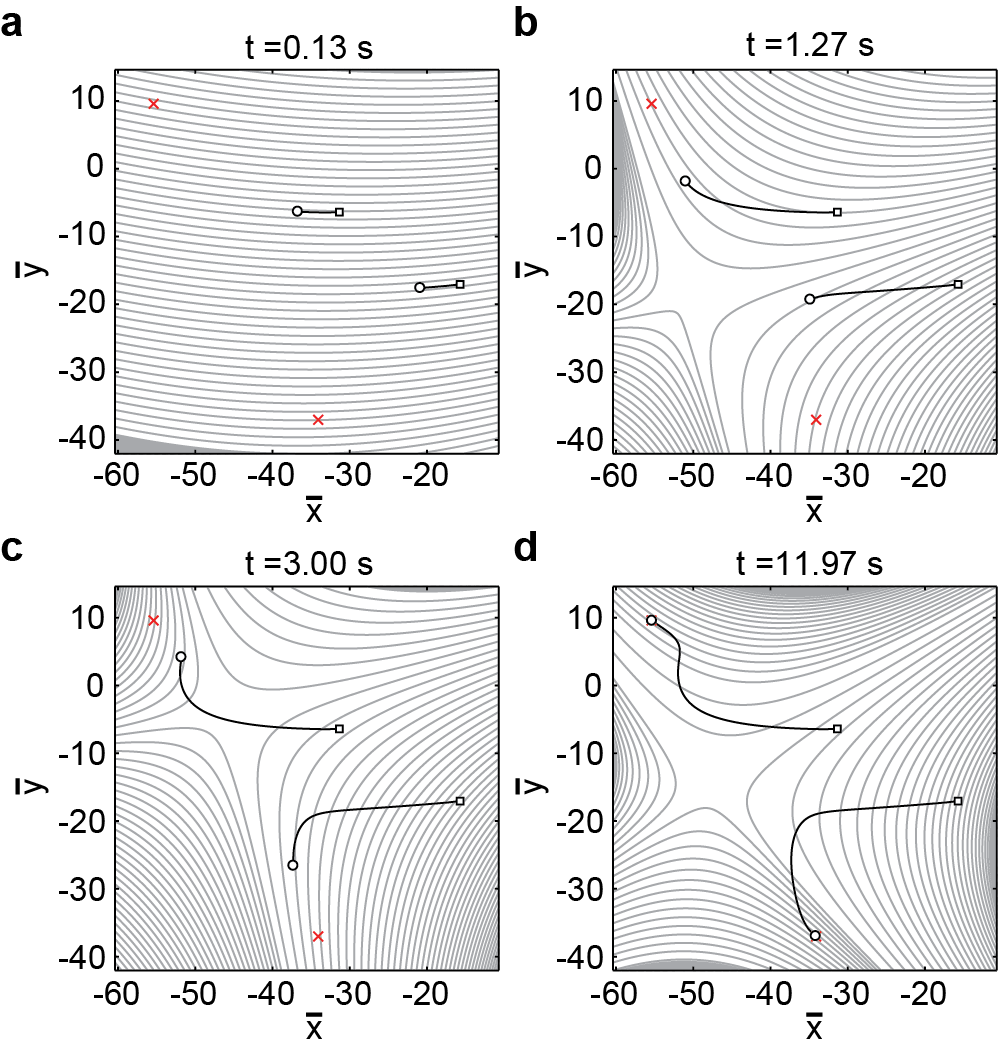}
\caption{\label{fig:diverge_streamlines} Streamlines at different instants in time during the manipulation process as particles are brought closer to each other, corresponding to the trapping process shown in \cref{fig:diverge_trajectories}. Streamlines shown at (a) t = 0.13 s (b) t = 1.27 s (c) t = 3.00 s (d) t = 11.97 s.}
\end{center}
\end{figure}

\begin{figure*}
\begin{center}
\includegraphics{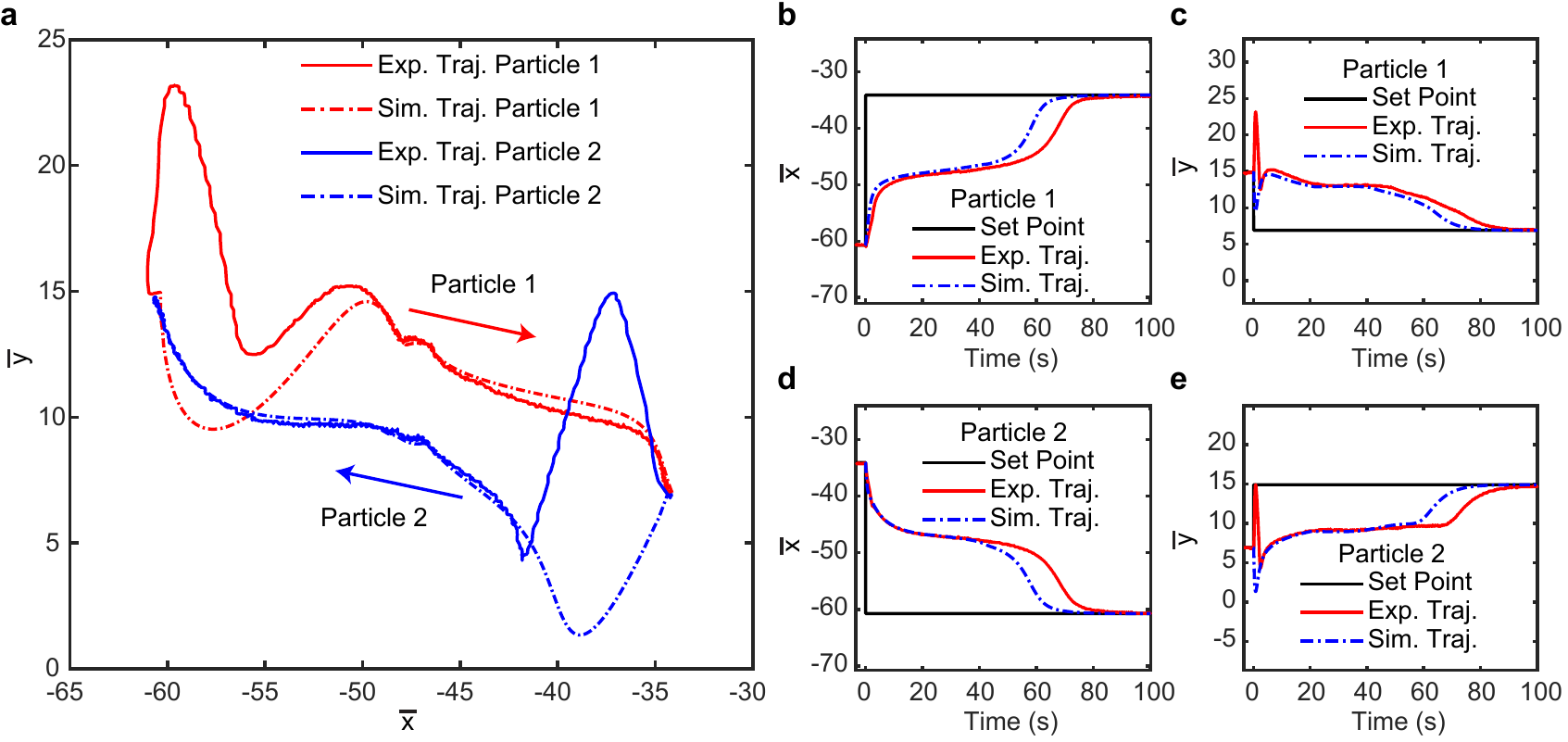}
\caption{\label{fig:interchange_trajectories} Particle trajectories for the case of interchanging the positions of two particles (corresponding to the schematic in \cref{fig:manipulation_experiments}c). Two particles are first trapped and then at time $t$=0, the trapping point for both particles is changed to switch their positions. (a) Particle trajectories from experiments and simulations where the arrows are drawn to show the overall direction of movement of each particle. (b) $x$-trajectory of particle 1 from experiments and simulations. (b) $y$-trajectory of particle 1 from experiments and simulations. (d) $x$-trajectory of particle 2 from experiments and simulations. (e) $y$-trajectory of particle 2 from experiments and simulations.}
\end{center}
\end{figure*}

\begin{figure}
\begin{center}
\includegraphics{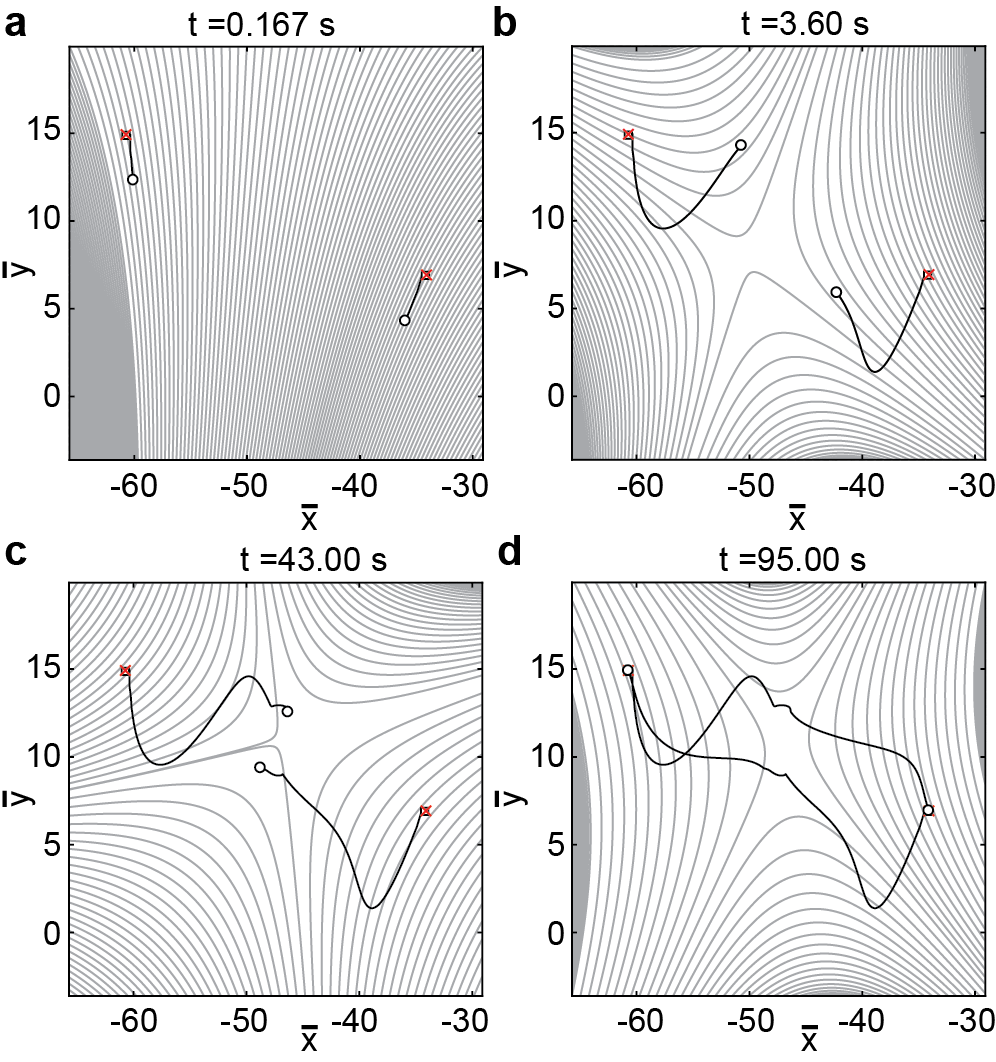}
\caption{\label{fig:interchange_streamlines} Streamlines at different instants of time during the trapping process as particles are brought away from each other (a) t = 0.167 s (b) t = 3.60 s (c) t = 43.00 s (d) t = 95.00 s}
\end{center}
\end{figure}

We further compare the experimentally applied flow rates, experimentally measured (actual) flow rates, and simulated flow rates for the manipulation process in \cref{fig:converge_flowrates}. The experimentally applied flow rates correspond to flow rates calculated using \cref{eq:pressure_to_flow}, whereas the actual flow rates are determined by using the experimentally measured pressure values $P_i$ (determined in real-time for each regulator), calculating $P_0$ using the relation $P_0 = \sum^6_{i=1}{P_i}/6$, and then back-calculating $q_i$ using \cref{eq:pressure_to_flow}. For completeness, the applied and measured pressures for each regulator are shown in \cref{fig:converge_pressures}. In general, we observe that deviations in the experimental and simulated particle trajectories (as observed in the $x$ and $y$ components) are correlated to differences in the applied and measured pressures. The pressure regulators generally show slightly different response times depending on whether a positive or negative pressure differential is applied. In particular, pressurizing tends to occur faster than depressurizing. Accordingly, for $P_1$ and $P_2$ in \cref{fig:converge_pressures}, the measured pressures tend to lag behind the applied pressures because initially a large negative pressure differential is required. However, the applied and measured pressures track fairly closely for the other regulators. This asymmetric difference in response rates leads to an initial transient deviation in the applied and actual flow rates, thereby resulting in the corresponding deviations in initial trajectories.

Given the generally good agreement between the experiments and simulations for particle trajectories and flow rates, we next used flow rates determined from simulations to analyze flow topologies. In this way, we plot streamlines during particle manipulation events to visualize the flow topologies (\cref{fig:converge_streamlines}). During the particle manipulation event corresponding to the trajectories in \cref{fig:converge_trajectories}, the particles first move in a roughly unidirectional flow in the same direction (\cref{fig:converge_streamlines}a). Next, a single stagnation point is created between the two particles which moves closer and closer to the centroid of the two particle positions (\cref{fig:converge_streamlines}b-c). Finally, both particles move towards their target positions driven by planar extensional flow, with a single stagnation point present between them (\cref{fig:converge_streamlines}d). In this way, the particles approach each other at a very close distance along the compressional axis (Supplementary Material Video 2). 

We next characterized flow topologies for the case of particles moving away from each other, which corresponds to the experiment shown schematically in \cref{fig:manipulation_experiments}b. Particle trajectories for this case are shown in \cref{fig:diverge_trajectories}, where particle 1 is initially trapped at [-31, -6] and particle 2 is initially trapped at [-16, -17], and at time $t$=0, the particle's target locations are set to [-55, 10] and [-34, -37], respectively. Again we generally observe good agreement between experimental and simulated trajectories for both particles (Supplementary Material Video 3). The deviations of experimental trajectories from the simulated trajectories are present only during initial transients, after which there appears to be good agreement. As in the prior case of particle motion during close approach, deviations are again caused by the slow depressurization response of the pressure regulators (Supplementary Material Fig. S1 and Fig. S2). 

We also characterized the evolution of the streamlines for this process (\cref{fig:diverge_streamlines}). Here, we observe that the particles are again initially transported in a unidirectional flow in one direction (\cref{fig:diverge_streamlines}a), after which a single stagnation point is created in the vicinity of the two particles, and both particles then move along planar extensional flow streamlines towards their final target (\cref{fig:diverge_streamlines}b-d). Interestingly, the trajectories of the particles are not reversed versions of the trajectories observed in \cref{fig:converge_trajectories}a. As seen previously, the choice of the objective function causes the particles first to be moved by a roughly uniform flow, which then transitions to a single stagnation point flow, thus leading to different trajectories when the particles are moved closer together rather than further apart (Supplementary Material Video 4). 

Finally, we characterized flow topologies corresponding to an interchange of particle positions, corresponding to the manipulation event schematically in \cref{fig:manipulation_experiments}c. Particle trajectories for experiments and simulations are shown in \cref{fig:interchange_trajectories}. In this experiment, particle 1 is initially at [-61, 15] and particle 2 is at [-34, 7], and at time $t$=0, the set points for both particles are interchanged. This manipulation event results in a large change in the particles' locations, so it is accompanied by correspondingly large values of flow rates and applied pressures (Supplementary Material Fig. S3 and Fig. S4). As a result, we initially observe deviations in the experimental and simulation trajectories for each particle (Supplementary Material Video 5). However, after the initial transients decay, the experimental and simulation trajectories generally show good agreement. In this process, a stagnation point is created between the two particles (\cref{fig:interchange_trajectories}b), and as the particles approach each other, the principal axes of the stagnation point flow rotate as the particles attempt to move past each other slowly (Supplementary Material Video 6). Comparing the simulation and the experimental trajectories near the instant of close approach, we posit that hydrodynamic interactions (HI) between the two particles may cause the simulation trajectory to approach the set point faster than the experimental trajectory, as particle-induced HI is not included in the simulations. Finally, once the particles are able to move away from each other, the remainder of the motion towards the final target location is fairly rapid, with the stagnation point rotating further to facilitate movement. In this experiment, the particles tend to approach each other closely approximately midway through the trapping process, which may result in localization issues as the tracking algorithm could confuse the identities of the two particles. To avoid this issue, we added an additional constraint to the MPC formulation such that particles cannot approach closer than 4 particle diameters.

In summary, flow topologies corresponding to the three canonical two-particle manipulation experiments consistently involve flow fields with zero stagnation points (for rapid repositioning) and one stagnation point (for adjusting and approaching final positions), rather than two stagnation points located at the center-of-mass of each particle. We conjecture that this set of flow topologies allows for more efficient, non-trivial dynamical manipulation of particles rather than a simple quasi-static displacement of trapping points. In general, the control algorithm tries to impose a unidirectional flow which moves the particles closer to their target locations, followed by switching to a stagnation point flow, where the algorithm is capable of both translating the stagnation point as well as rotating the principal axes of the extensional flow. 

\subsection{Effect of the controller weights.}
We now focus on the impact of the controller parameters on the speed of manipulation, including the effect of the controller weight $\beta$ associated with the flow rates and the controller weight $\gamma$ associated with the end penalty term $\gamma$. The optimal trajectories sensitively depend on the initial and final positions of the particles. In general, we can divide these trajectories into two categories, including one case wherein the distance of each particle to the set point smoothly decays monotonically and a second case wherein the distance of a particle initially decays to zero, then increases, and finally decays to 0 again. We consider the effect of the controller weights on both cases in order to fully characterize the sensitivity to speed of manipulation. In all cases, the default values for the controller weights are $\beta=10^{-4}$ and $\gamma=1000$. We characterize the speed of manipulation by measuring the time taken by each particle to first reach within 1 particle diameter of its target location. 
\begin{figure}
\begin{center}
\includegraphics{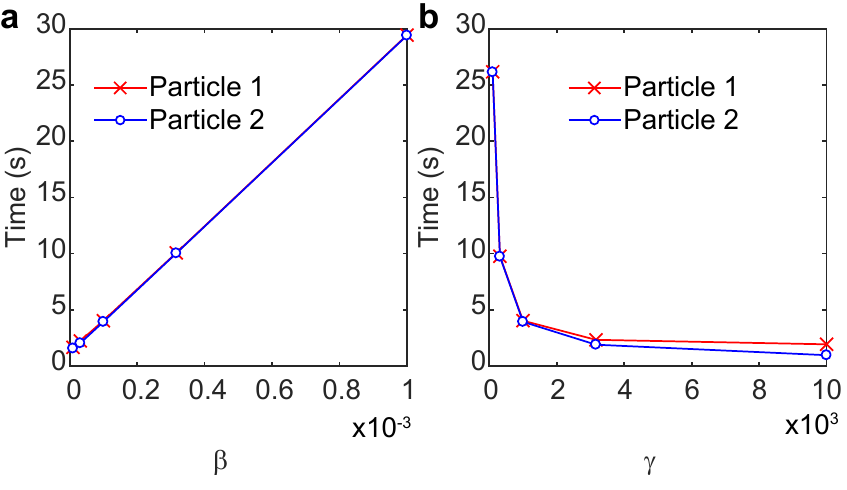}
\caption{\label{fig:effect_control_wts_smooth} Effect of the controller weights $\beta$ and $\gamma$ on the time required for two particles to reach their final target position. (a) Effect of the controller weight $\beta$ on the time required to reach 1 particle diameter. (b) Effect of end penalty term weight $\gamma$ on the time required to reach 1 particle diameter.}
\end{center}
\end{figure}

We began by considering the case where both particles' distance to the target position smoothly decays to 0 nearly monotonically. We first vary the control weight $\beta$ while holding $\gamma$ constant. Particle 1 is trapped at [10,10], and particle 2 is trapped at [-10,-10]. At time $t$=0, the set points for these particles are changed to [-30,10] and [-10,-30], respectively. Five different values of $\beta = 1.0\times10^{-5},3.2\times10^{-4},1.0\times10^{-4},3.2\times10^{-3},1.0\times10^{-3}$ are selected, which are spaced equally on a logarithmic scale between $10^{-5}$ and $10^{-3}$. 

\cref{fig:effect_control_wts_smooth}a shows the time required for particles to reach within 1 particle diameter for a given value of $\beta$. In this case, the distances of both particles to the target position decay smoothly to 0 roughly monotonically (Fig. S5a,b). Interestingly, a linear relationship between the value of $\beta$ and the time required for manipulation is observed. $\beta$ is a regularizer for the flow rate applied during manipulation, so small values of $\beta$ allow for larger values of the flow rates for the same value of the objective function. As a result, the time required to reach the target location increases as $\beta$ increases, because the flow rates are restricted from increasing without limit. In addition, even though both particles traverse different distances along their trajectories, they require roughly the same amount of time to reach their target location. This behavior occurs because the deviation of each particle from its target location is equally weighted in the objective function, so the flow rates obtained by minimizing the objective function are such that they cause both particles to arrive at the target in the same time. Moreover, we observe that as $\beta$ decreases, the path begins to approach a straight line between the initial and final position (Fig. S5c).

We next varied the control parameter $\gamma$ corresponding to the end penalty term. We select five different values of $\gamma = 1.0\times10^{2},3.2\times10^{2},1.0\times10^{3},3.2\times10^{3},1.0\times10^{4}$, which are again equally spaced on a logarithmic scale between $10^{2}$ and $10^{4}$. \cref{fig:effect_control_wts_smooth}b shows the effect of $\gamma$ on the time taken by each particle to reach within 1 particle diameter of the target location. We again observe that both particles require nearly the same amount of time to reach their respective target locations. As previously discussed, $\gamma$ plays a critical role in removing offsets at the end of the prediction horizon. When the value of $\gamma$ is increased, the deviation of particle position at the end of the prediction horizon becomes more important compared to the flow rates and particle deviations over the prediction horizon. For large values of $\gamma$, the controller chooses larger flow rates to reduce the offset in particle position in an attempt to minimize the contribution of the end penalty term. In this way, increasing $\gamma$ leads to a reduction in the time required to reach the target location, which is shown in \cref{fig:effect_control_wts_smooth}b. In addition, increasing the value of $\gamma$ beyond a certain value appears to have a negligible effect on the time required to reach the target location (\cref{fig:effect_control_wts_smooth}b). Particle trajectories corresponding to these simulations are shown in Fig. S6c, while the individual particle distances over time are shown in Fig. S6a,b. We observe that the initial curved trajectory for both particles begins to straighten out with larger values of $\gamma$, which suggests that the particles are approaching the straight line trajectory between their initial and desired final locations. We characterized behavior for multiple initial and end point positions that likewise generated smooth trajectories. In all cases, we observe a similar dependence of $\beta$ and $\gamma$ on the time required for particles to reach final position (Figs. S7-S12).

\begin{figure}
\begin{center}
\includegraphics{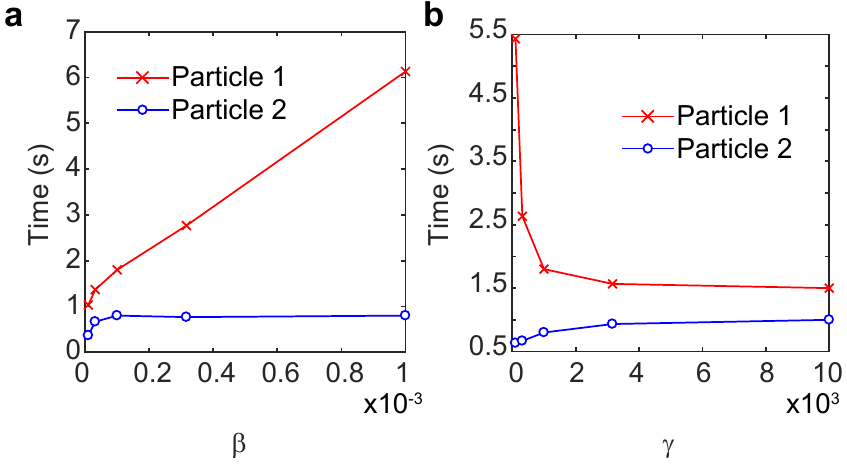}
\caption{\label{fig:effect_control_wts_nonsmooth} Effect of the controller weights $\beta$ and $\gamma$ on the time required for two particles to reach their final target position when a non-smooth trajectory is followed. (a) Effect of the controller weight $\beta$ on the time required to reach 1 particle diameter. (b) Effect of end penalty term weight $\gamma$ on the time required to reach 1 particle diameter.}
\end{center}
\end{figure}

Next, we consider the case where the distance of either particle to the target location does not smoothly decay to zero. To probe this case, we initially trap two particles at [20,30] and [-50,-40] and then change their set points to [30, 40] and [-30,-20], respectively. \cref{fig:effect_control_wts_nonsmooth}a shows the effect of variation of $\beta$ on the speed of manipulation. In this case, the trajectory for particle 2 was non-smooth, as the distance to the set point did not monotonically decay to zero (Fig. S13b), whereas the distance of particle 1 smoothly decays to zero (Fig. S13a). Accordingly, we find that the time required by particle 1 to reach the final position increases with $\beta$, but for particle 2, the time first increases and then saturates (\cref{fig:effect_control_wts_nonsmooth}a). Interestingly, both particles require different times to reach within a particle diameter of the target position, in contrast to the smooth trajectory case. \cref{fig:effect_control_wts_nonsmooth}b shows the effect of variation of $\gamma$ on the time required to reach 1 particle diameter. Particle 1 has a familiar trend with $\gamma$ as the time decreases with increasing gamma initially, and then has negligible changes for larger values of $\gamma$. On the other hand, particle 2's time increases initially with $\gamma$, and then saturates. Again, these results suggest that particle 1 has a smoothly decaying distance to the target location, whereas particle 2 has a non-smooth trajectory (Fig. S14a,b). 

\section{Conclusions}
In this work, we systematically analyze flow topologies during multiplexed particle manipulation using a Stokes trap. In MPC, the controller determines optimal flow rates to minimize an objective function. Compared to the initial version of the automated hydrodynamic trap \cite{tanyeri_manipulation_2013}, which only has a simplistic design with a single stagnation point, the flow topologies generated in the Stokes trap are complex and time-dependent. Achieving an understanding of the dynamic flow fields during particle trapping will be essential for future applications in soft matter and flow-induced particle deformation experiments using this technique.  

Our results show that a six-channel cross-slot is capable of three distinct flow topologies with zero, one, and two stagnation points, which can be accessed by changing the flow rates in each channel. Using a combination of experiments and simulations, we find that optimal particle manipulation involves a two-stage strategy: in the initial stage, there are no stagnation points in the vicinity of the two particles, however, in a later stage, a single stagnation point is generated between the two particles. The action of the controller is then to translate this single stagnation point and to rotate the orientation of the principal axes of extension and compression to enable a precise, exponential approach to the desired particle positions. In this way, a simpler flow field topology with zero or one stagnation point is superior in performing particle manipulation as opposed to a more complicated flow pattern, where two stagnation points could potentially control particle position.

The apparent advantage of zero or one stagnation points for manipulating two particles manifests itself in both shorter duration of the relocation process and in smaller flow rates needed to perform the task, both of which are crucial limitations in practical applications. Our findings have significant consequences for particle manipulation in general, but in particular for applications that involve particle-particle interactions. For instance, if two vesicles are trapped in this setup, and they are brought close to each other in a controlled fashion for adhesion experiments, our work establishes that both the particles essentially experience planar extensional flow, instead of the complicated two stagnation point flow (as shown in \cref{fig:three_topologies}b). 

We further explore the impact of the controller parameters on the speed of manipulation using simulations. In general, our results show that the time required to reach the target location is the same for both the trapped particles when they follow smooth trajectories. In addition, we found that increasing $\gamma$ leads to faster manipulation initially for smooth trajectories, after which the performance saturates. Broadly, this work provides a systematic characterization of the performance of the Stokes trap by examining the evolution of flow topologies during multiplexed particle manipulation, and further shows how the controller parameters can be tuned to obtain the desired level of trapping performance. Taken together, these results will help guide future experimental studies on deformable soft materials in well defined flow fields using a Stokes trap.

\section*{Author contributions}
A.S., S.H. and C.M.S conceived the project and designed the experiments. A.S. performed the experiments, while A.S. and D.K. analyzed the experimental data. S.H. and C.M.S supervised the research and all authors wrote the manuscript.  

\begin{acknowledgments}
The authors thank Prof. Christopher V. Rao for insightful discussions. This work was supported by the National Science Foundation by Award \# NSF CBET 1704668.
\end{acknowledgments}

% \verb|end{document}|
% \bibliography{refs}

\begin{thebibliography}{33}%
\makeatletter
\providecommand \@ifxundefined [1]{%
 \@ifx{#1\undefined}
}%
\providecommand \@ifnum [1]{%
 \ifnum #1\expandafter \@firstoftwo
 \else \expandafter \@secondoftwo
 \fi
}%
\providecommand \@ifx [1]{%
 \ifx #1\expandafter \@firstoftwo
 \else \expandafter \@secondoftwo
 \fi
}%
\providecommand \natexlab [1]{#1}%
\providecommand \enquote  [1]{``#1''}%
\providecommand \bibnamefont  [1]{#1}%
\providecommand \bibfnamefont [1]{#1}%
\providecommand \citenamefont [1]{#1}%
\providecommand \href@noop [0]{\@secondoftwo}%
\providecommand \href [0]{\begingroup \@sanitize@url \@href}%
\providecommand \@href[1]{\@@startlink{#1}\@@href}%
\providecommand \@@href[1]{\endgroup#1\@@endlink}%
\providecommand \@sanitize@url [0]{\catcode `\\12\catcode `\$12\catcode
  `\&12\catcode `\#12\catcode `\^12\catcode `\_12\catcode `\%12\relax}%
\providecommand \@@startlink[1]{}%
\providecommand \@@endlink[0]{}%
\providecommand \url  [0]{\begingroup\@sanitize@url \@url }%
\providecommand \@url [1]{\endgroup\@href {#1}{\urlprefix }}%
\providecommand \urlprefix  [0]{URL }%
\providecommand \Eprint [0]{\href }%
\providecommand \doibase [0]{https://doi.org/}%
\providecommand \selectlanguage [0]{\@gobble}%
\providecommand \bibinfo  [0]{\@secondoftwo}%
\providecommand \bibfield  [0]{\@secondoftwo}%
\providecommand \translation [1]{[#1]}%
\providecommand \BibitemOpen [0]{}%
\providecommand \bibitemStop [0]{}%
\providecommand \bibitemNoStop [0]{.\EOS\space}%
\providecommand \EOS [0]{\spacefactor3000\relax}%
\providecommand \BibitemShut  [1]{\csname bibitem#1\endcsname}%
\let\auto@bib@innerbib\@empty
%</preamble>
\bibitem [{\citenamefont {Chiou}\ \emph {et~al.}(2005)\citenamefont {Chiou},
  \citenamefont {Ohta},\ and\ \citenamefont {Wu}}]{chiou_massively_2005}%
  \BibitemOpen
  \bibfield  {author} {\bibinfo {author} {\bibfnamefont {P.~Y.}\ \bibnamefont
  {Chiou}}, \bibinfo {author} {\bibfnamefont {A.~T.}\ \bibnamefont {Ohta}},\
  and\ \bibinfo {author} {\bibfnamefont {M.~C.}\ \bibnamefont {Wu}},\
  }\bibfield  {title} {\bibinfo {title} {Massively parallel manipulation of
  single cells and microparticles using optical images},\ }\href@noop {}
  {\bibfield  {journal} {\bibinfo  {journal} {Nature}\ }\textbf {\bibinfo
  {volume} {436}},\ \bibinfo {pages} {370} (\bibinfo {year}
  {2005})}\BibitemShut {NoStop}%
\bibitem [{\citenamefont {Wang}\ \emph {et~al.}(1997)\citenamefont {Wang},
  \citenamefont {Yin}, \citenamefont {Landick}, \citenamefont {Gelles},\ and\
  \citenamefont {Block}}]{wang_stretching_1997}%
  \BibitemOpen
  \bibfield  {author} {\bibinfo {author} {\bibfnamefont {M.~D.}\ \bibnamefont
  {Wang}}, \bibinfo {author} {\bibfnamefont {H.}~\bibnamefont {Yin}}, \bibinfo
  {author} {\bibfnamefont {R.}~\bibnamefont {Landick}}, \bibinfo {author}
  {\bibfnamefont {J.}~\bibnamefont {Gelles}},\ and\ \bibinfo {author}
  {\bibfnamefont {S.~M.}\ \bibnamefont {Block}},\ }\bibfield  {title} {\bibinfo
  {title} {Stretching {DNA} with optical tweezers},\ }\href@noop {} {\bibfield
  {journal} {\bibinfo  {journal} {Biophys. J.}\ }\textbf {\bibinfo {volume}
  {72}},\ \bibinfo {pages} {1335} (\bibinfo {year} {1997})}\BibitemShut
  {NoStop}%
\bibitem [{\citenamefont {Chemla}\ \emph {et~al.}(2005)\citenamefont {Chemla},
  \citenamefont {Aathavan}, \citenamefont {Michaelis}, \citenamefont {Grimes},
  \citenamefont {Jardine}, \citenamefont {Anderson},\ and\ \citenamefont
  {Bustamante}}]{chemla_mechanism_2005}%
  \BibitemOpen
  \bibfield  {author} {\bibinfo {author} {\bibfnamefont {Y.~R.}\ \bibnamefont
  {Chemla}}, \bibinfo {author} {\bibfnamefont {K.}~\bibnamefont {Aathavan}},
  \bibinfo {author} {\bibfnamefont {J.}~\bibnamefont {Michaelis}}, \bibinfo
  {author} {\bibfnamefont {S.}~\bibnamefont {Grimes}}, \bibinfo {author}
  {\bibfnamefont {P.~J.}\ \bibnamefont {Jardine}}, \bibinfo {author}
  {\bibfnamefont {D.~L.}\ \bibnamefont {Anderson}},\ and\ \bibinfo {author}
  {\bibfnamefont {C.}~\bibnamefont {Bustamante}},\ }\bibfield  {title}
  {\bibinfo {title} {Mechanism of force generation of a viral {DNA} packaging
  motor},\ }\href@noop {} {\bibfield  {journal} {\bibinfo  {journal} {Cell}\
  }\textbf {\bibinfo {volume} {122}},\ \bibinfo {pages} {683} (\bibinfo {year}
  {2005})}\BibitemShut {NoStop}%
\bibitem [{\citenamefont {Ashkin}\ \emph {et~al.}(1986)\citenamefont {Ashkin},
  \citenamefont {Dziedzic}, \citenamefont {Bjorkholm},\ and\ \citenamefont
  {Chu}}]{ashkin_observation_1986}%
  \BibitemOpen
  \bibfield  {author} {\bibinfo {author} {\bibfnamefont {A.}~\bibnamefont
  {Ashkin}}, \bibinfo {author} {\bibfnamefont {J.~M.}\ \bibnamefont
  {Dziedzic}}, \bibinfo {author} {\bibfnamefont {J.~E.}\ \bibnamefont
  {Bjorkholm}},\ and\ \bibinfo {author} {\bibfnamefont {S.}~\bibnamefont
  {Chu}},\ }\bibfield  {title} {\bibinfo {title} {Observation of a single-beam
  gradient force optical trap for dielectric particles},\ }\href@noop {}
  {\bibfield  {journal} {\bibinfo  {journal} {Opt. Lett.}\ }\textbf {\bibinfo
  {volume} {11}},\ \bibinfo {pages} {288} (\bibinfo {year} {1986})}\BibitemShut
  {NoStop}%
\bibitem [{\citenamefont {Grier}(2003)}]{grier_revolution_2003}%
  \BibitemOpen
  \bibfield  {author} {\bibinfo {author} {\bibfnamefont {D.~G.}\ \bibnamefont
  {Grier}},\ }\bibfield  {title} {\bibinfo {title} {A revolution in optical
  manipulation},\ }\href@noop {} {\bibfield  {journal} {\bibinfo  {journal}
  {Nature}\ }\textbf {\bibinfo {volume} {424}},\ \bibinfo {pages} {810}
  (\bibinfo {year} {2003})}\BibitemShut {NoStop}%
\bibitem [{\citenamefont {Gosse}\ and\ \citenamefont
  {Croquette}(2002)}]{gosse_magnetic_2002}%
  \BibitemOpen
  \bibfield  {author} {\bibinfo {author} {\bibfnamefont {C.}~\bibnamefont
  {Gosse}}\ and\ \bibinfo {author} {\bibfnamefont {V.}~\bibnamefont
  {Croquette}},\ }\bibfield  {title} {\bibinfo {title} {Magnetic tweezers:
  micromanipulation and force measurement at the molecular level},\ }\href@noop
  {} {\bibfield  {journal} {\bibinfo  {journal} {Biophys. J.}\ }\textbf
  {\bibinfo {volume} {82}},\ \bibinfo {pages} {3314} (\bibinfo {year}
  {2002})}\BibitemShut {NoStop}%
\bibitem [{\citenamefont {Sarkar}\ and\ \citenamefont
  {Rybenkov}(2016)}]{sarkar_guide_2016}%
  \BibitemOpen
  \bibfield  {author} {\bibinfo {author} {\bibfnamefont {R.}~\bibnamefont
  {Sarkar}}\ and\ \bibinfo {author} {\bibfnamefont {V.~V.}\ \bibnamefont
  {Rybenkov}},\ }\bibfield  {title} {\bibinfo {title} {A guide to magnetic
  tweezers and their applications},\ }\href@noop {} {\bibfield  {journal}
  {\bibinfo  {journal} {Front. Phys.}\ }\textbf {\bibinfo {volume} {4}},\
  \bibinfo {pages} {48} (\bibinfo {year} {2016})}\BibitemShut {NoStop}%
\bibitem [{\citenamefont {Cohen}\ and\ \citenamefont
  {Moerner}(2006)}]{cohen_suppressing_2006}%
  \BibitemOpen
  \bibfield  {author} {\bibinfo {author} {\bibfnamefont {A.~E.}\ \bibnamefont
  {Cohen}}\ and\ \bibinfo {author} {\bibfnamefont {W.~E.}\ \bibnamefont
  {Moerner}},\ }\bibfield  {title} {\bibinfo {title} {Suppressing {Brownian}
  motion of individual biomolecules in solution},\ }\href@noop {} {\bibfield
  {journal} {\bibinfo  {journal} {Proc. Natl. Acad. Sci. U.S.A.}\ }\textbf
  {\bibinfo {volume} {103}},\ \bibinfo {pages} {4362} (\bibinfo {year}
  {2006})}\BibitemShut {NoStop}%
\bibitem [{\citenamefont {Ropp}\ \emph {et~al.}(2010)\citenamefont {Ropp},
  \citenamefont {Probst}, \citenamefont {Cummins}, \citenamefont {Kumar},
  \citenamefont {Berglund}, \citenamefont {Raghavan}, \citenamefont {Waks},\
  and\ \citenamefont {Shapiro}}]{ropp_manipulating_2010}%
  \BibitemOpen
  \bibfield  {author} {\bibinfo {author} {\bibfnamefont {C.}~\bibnamefont
  {Ropp}}, \bibinfo {author} {\bibfnamefont {R.}~\bibnamefont {Probst}},
  \bibinfo {author} {\bibfnamefont {Z.}~\bibnamefont {Cummins}}, \bibinfo
  {author} {\bibfnamefont {R.}~\bibnamefont {Kumar}}, \bibinfo {author}
  {\bibfnamefont {A.~J.}\ \bibnamefont {Berglund}}, \bibinfo {author}
  {\bibfnamefont {S.~R.}\ \bibnamefont {Raghavan}}, \bibinfo {author}
  {\bibfnamefont {E.}~\bibnamefont {Waks}},\ and\ \bibinfo {author}
  {\bibfnamefont {B.}~\bibnamefont {Shapiro}},\ }\bibfield  {title} {\bibinfo
  {title} {Manipulating quantum dots to nanometer precision by control of
  flow},\ }\href@noop {} {\bibfield  {journal} {\bibinfo  {journal} {Nano
  Lett.}\ }\textbf {\bibinfo {volume} {10}},\ \bibinfo {pages} {2525} (\bibinfo
  {year} {2010})}\BibitemShut {NoStop}%
\bibitem [{\citenamefont {Hertz}(1995)}]{hertz_standing-wave_1995}%
  \BibitemOpen
  \bibfield  {author} {\bibinfo {author} {\bibfnamefont {H.~M.}\ \bibnamefont
  {Hertz}},\ }\bibfield  {title} {\bibinfo {title} {Standing-wave acoustic trap
  for nonintrusive positioning of microparticles},\ }\href@noop {} {\bibfield
  {journal} {\bibinfo  {journal} {J. Appl. Phys.}\ }\textbf {\bibinfo {volume}
  {78}},\ \bibinfo {pages} {4845} (\bibinfo {year} {1995})}\BibitemShut
  {NoStop}%
\bibitem [{\citenamefont {Guo}\ \emph {et~al.}(2016)\citenamefont {Guo},
  \citenamefont {Mao}, \citenamefont {Chen}, \citenamefont {Xie}, \citenamefont
  {Lata}, \citenamefont {Li}, \citenamefont {Ren}, \citenamefont {Liu},
  \citenamefont {Yang},\ and\ \citenamefont
  {Dao}}]{guo_three-dimensional_2016}%
  \BibitemOpen
  \bibfield  {author} {\bibinfo {author} {\bibfnamefont {F.}~\bibnamefont
  {Guo}}, \bibinfo {author} {\bibfnamefont {Z.}~\bibnamefont {Mao}}, \bibinfo
  {author} {\bibfnamefont {Y.}~\bibnamefont {Chen}}, \bibinfo {author}
  {\bibfnamefont {Z.}~\bibnamefont {Xie}}, \bibinfo {author} {\bibfnamefont
  {J.~P.}\ \bibnamefont {Lata}}, \bibinfo {author} {\bibfnamefont
  {P.}~\bibnamefont {Li}}, \bibinfo {author} {\bibfnamefont {L.}~\bibnamefont
  {Ren}}, \bibinfo {author} {\bibfnamefont {J.}~\bibnamefont {Liu}}, \bibinfo
  {author} {\bibfnamefont {J.}~\bibnamefont {Yang}},\ and\ \bibinfo {author}
  {\bibfnamefont {M.}~\bibnamefont {Dao}},\ }\bibfield  {title} {\bibinfo
  {title} {Three-dimensional manipulation of single cells using surface
  acoustic waves},\ }\href@noop {} {\bibfield  {journal} {\bibinfo  {journal}
  {Proc. Natl. Acad. Sci. U.S.A.}\ }\textbf {\bibinfo {volume} {113}},\
  \bibinfo {pages} {1522} (\bibinfo {year} {2016})}\BibitemShut {NoStop}%
\bibitem [{\citenamefont {Petit}\ \emph {et~al.}(2011)\citenamefont {Petit},
  \citenamefont {Zhang}, \citenamefont {Peyer}, \citenamefont {Kratochvil},\
  and\ \citenamefont {Nelson}}]{petit_selective_2011}%
  \BibitemOpen
  \bibfield  {author} {\bibinfo {author} {\bibfnamefont {T.}~\bibnamefont
  {Petit}}, \bibinfo {author} {\bibfnamefont {L.}~\bibnamefont {Zhang}},
  \bibinfo {author} {\bibfnamefont {K.~E.}\ \bibnamefont {Peyer}}, \bibinfo
  {author} {\bibfnamefont {B.~E.}\ \bibnamefont {Kratochvil}},\ and\ \bibinfo
  {author} {\bibfnamefont {B.~J.}\ \bibnamefont {Nelson}},\ }\bibfield  {title}
  {\bibinfo {title} {Selective trapping and manipulation of microscale objects
  using mobile microvortices},\ }\href@noop {} {\bibfield  {journal} {\bibinfo
  {journal} {Nano Lett.}\ }\textbf {\bibinfo {volume} {12}},\ \bibinfo {pages}
  {156} (\bibinfo {year} {2011})}\BibitemShut {NoStop}%
\bibitem [{\citenamefont {Tanyeri}\ \emph {et~al.}(2011)\citenamefont
  {Tanyeri}, \citenamefont {Ranka}, \citenamefont {Sittipolkul},\ and\
  \citenamefont {Schroeder}}]{tanyeri_microfluidic-based_2011}%
  \BibitemOpen
  \bibfield  {author} {\bibinfo {author} {\bibfnamefont {M.}~\bibnamefont
  {Tanyeri}}, \bibinfo {author} {\bibfnamefont {M.}~\bibnamefont {Ranka}},
  \bibinfo {author} {\bibfnamefont {N.}~\bibnamefont {Sittipolkul}},\ and\
  \bibinfo {author} {\bibfnamefont {C.~M.}\ \bibnamefont {Schroeder}},\
  }\bibfield  {title} {\bibinfo {title} {A microfluidic-based hydrodynamic
  trap: design and implementation},\ }\href@noop {} {\bibfield  {journal}
  {\bibinfo  {journal} {Lab. Chip}\ }\textbf {\bibinfo {volume} {11}},\
  \bibinfo {pages} {1786} (\bibinfo {year} {2011})}\BibitemShut {NoStop}%
\bibitem [{\citenamefont {Tanyeri}\ and\ \citenamefont
  {Schroeder}(2013)}]{tanyeri_manipulation_2013}%
  \BibitemOpen
  \bibfield  {author} {\bibinfo {author} {\bibfnamefont {M.}~\bibnamefont
  {Tanyeri}}\ and\ \bibinfo {author} {\bibfnamefont {C.~M.}\ \bibnamefont
  {Schroeder}},\ }\bibfield  {title} {\bibinfo {title} {Manipulation and
  confinement of single particles using fluid flow},\ }\href@noop {} {\bibfield
   {journal} {\bibinfo  {journal} {Nano Lett.}\ }\textbf {\bibinfo {volume}
  {13}},\ \bibinfo {pages} {2357} (\bibinfo {year} {2013})}\BibitemShut
  {NoStop}%
\bibitem [{\citenamefont {Shenoy}\ \emph {et~al.}(2015)\citenamefont {Shenoy},
  \citenamefont {Tanyeri},\ and\ \citenamefont
  {Schroeder}}]{shenoy_characterizing_2015}%
  \BibitemOpen
  \bibfield  {author} {\bibinfo {author} {\bibfnamefont {A.}~\bibnamefont
  {Shenoy}}, \bibinfo {author} {\bibfnamefont {M.}~\bibnamefont {Tanyeri}},\
  and\ \bibinfo {author} {\bibfnamefont {C.~M.}\ \bibnamefont {Schroeder}},\
  }\bibfield  {title} {\bibinfo {title} {Characterizing the performance of the
  hydrodynamic trap using a control-based approach},\ }\href@noop {} {\bibfield
   {journal} {\bibinfo  {journal} {Microfluid. Nanofluid.}\ }\textbf {\bibinfo
  {volume} {18}},\ \bibinfo {pages} {1055} (\bibinfo {year}
  {2015})}\BibitemShut {NoStop}%
\bibitem [{\citenamefont {Shenoy}\ \emph {et~al.}(2016)\citenamefont {Shenoy},
  \citenamefont {Rao},\ and\ \citenamefont {Schroeder}}]{shenoy_stokes_2016}%
  \BibitemOpen
  \bibfield  {author} {\bibinfo {author} {\bibfnamefont {A.}~\bibnamefont
  {Shenoy}}, \bibinfo {author} {\bibfnamefont {C.~V.}\ \bibnamefont {Rao}},\
  and\ \bibinfo {author} {\bibfnamefont {C.~M.}\ \bibnamefont {Schroeder}},\
  }\bibfield  {title} {\bibinfo {title} {Stokes trap for multiplexed particle
  manipulation and assembly using fluidics},\ }\href@noop {} {\bibfield
  {journal} {\bibinfo  {journal} {Proc. Natl. Acad. Sci. U.S.A.}\ }\textbf
  {\bibinfo {volume} {113}},\ \bibinfo {pages} {3976} (\bibinfo {year}
  {2016})}\BibitemShut {NoStop}%
\bibitem [{\citenamefont {Kumar}\ \emph {et~al.}(2019)\citenamefont {Kumar},
  \citenamefont {Shenoy}, \citenamefont {Li},\ and\ \citenamefont
  {Schroeder}}]{kumar_orientation_2019}%
  \BibitemOpen
  \bibfield  {author} {\bibinfo {author} {\bibfnamefont {D.}~\bibnamefont
  {Kumar}}, \bibinfo {author} {\bibfnamefont {A.}~\bibnamefont {Shenoy}},
  \bibinfo {author} {\bibfnamefont {S.}~\bibnamefont {Li}},\ and\ \bibinfo
  {author} {\bibfnamefont {C.~M.}\ \bibnamefont {Schroeder}},\ }\bibfield
  {title} {\bibinfo {title} {Orientation control and nonlinear trajectory
  tracking of colloidal particles using microfluidics},\ }\href
  {http://arxiv.org/abs/1907.08567} {\bibfield  {journal} {\bibinfo  {journal}
  {arXiv:1907.08567}\ } (\bibinfo {year} {2019})}\BibitemShut {NoStop}%
\bibitem [{\citenamefont {Taylor}(1934)}]{taylor_formation_1934}%
  \BibitemOpen
  \bibfield  {author} {\bibinfo {author} {\bibfnamefont {G.~I.}\ \bibnamefont
  {Taylor}},\ }\bibfield  {title} {\bibinfo {title} {The formation of emulsions
  in definable fields of flow},\ }\href@noop {} {\bibfield  {journal} {\bibinfo
   {journal} {Proc. Royal Soc. A}\ }\textbf {\bibinfo {volume} {146}},\
  \bibinfo {pages} {501} (\bibinfo {year} {1934})}\BibitemShut {NoStop}%
\bibitem [{\citenamefont {Bentley}\ and\ \citenamefont
  {Leal}(1986)}]{bentley_computer-controlled_1986}%
  \BibitemOpen
  \bibfield  {author} {\bibinfo {author} {\bibfnamefont {B.~J.}\ \bibnamefont
  {Bentley}}\ and\ \bibinfo {author} {\bibfnamefont {L.~G.}\ \bibnamefont
  {Leal}},\ }\bibfield  {title} {\bibinfo {title} {A computer-controlled
  four-roll mill for investigations of particle and drop dynamics in
  two-dimensional linear shear flows},\ }\href@noop {} {\bibfield  {journal}
  {\bibinfo  {journal} {J. Fluid Mech.}\ }\textbf {\bibinfo {volume} {167}},\
  \bibinfo {pages} {219} (\bibinfo {year} {1986})}\BibitemShut {NoStop}%
\bibitem [{\citenamefont {Hudson}\ \emph {et~al.}(2004)\citenamefont {Hudson},
  \citenamefont {Phelan~Jr}, \citenamefont {Handler}, \citenamefont {Cabral},
  \citenamefont {Migler},\ and\ \citenamefont
  {Amis}}]{hudson_microfluidic_2004}%
  \BibitemOpen
  \bibfield  {author} {\bibinfo {author} {\bibfnamefont {S.~D.}\ \bibnamefont
  {Hudson}}, \bibinfo {author} {\bibfnamefont {F.~R.}\ \bibnamefont
  {Phelan~Jr}}, \bibinfo {author} {\bibfnamefont {M.~D.}\ \bibnamefont
  {Handler}}, \bibinfo {author} {\bibfnamefont {J.~T.}\ \bibnamefont {Cabral}},
  \bibinfo {author} {\bibfnamefont {K.~B.}\ \bibnamefont {Migler}},\ and\
  \bibinfo {author} {\bibfnamefont {E.~J.}\ \bibnamefont {Amis}},\ }\bibfield
  {title} {\bibinfo {title} {Microfluidic analog of the four-roll mill},\
  }\href@noop {} {\bibfield  {journal} {\bibinfo  {journal} {Appl. Phys.
  Lett.}\ }\textbf {\bibinfo {volume} {85}},\ \bibinfo {pages} {335} (\bibinfo
  {year} {2004})}\BibitemShut {NoStop}%
\bibitem [{\citenamefont {Lee}\ \emph {et~al.}(2007)\citenamefont {Lee},
  \citenamefont {Dylla-Spears}, \citenamefont {Teclemariam},\ and\
  \citenamefont {Muller}}]{lee_microfluidic_2007}%
  \BibitemOpen
  \bibfield  {author} {\bibinfo {author} {\bibfnamefont {J.~S.}\ \bibnamefont
  {Lee}}, \bibinfo {author} {\bibfnamefont {R.}~\bibnamefont {Dylla-Spears}},
  \bibinfo {author} {\bibfnamefont {N.~P.}\ \bibnamefont {Teclemariam}},\ and\
  \bibinfo {author} {\bibfnamefont {S.~J.}\ \bibnamefont {Muller}},\ }\bibfield
   {title} {\bibinfo {title} {Microfluidic four-roll mill for all flow types},\
  }\href@noop {} {\bibfield  {journal} {\bibinfo  {journal} {Appl. Phys.
  Lett.}\ }\textbf {\bibinfo {volume} {90}},\ \bibinfo {pages} {074103}
  (\bibinfo {year} {2007})}\BibitemShut {NoStop}%
\bibitem [{\citenamefont {Deschamps}\ \emph {et~al.}(2009)\citenamefont
  {Deschamps}, \citenamefont {Kantsler}, \citenamefont {Segre},\ and\
  \citenamefont {Steinberg}}]{deschamps_dynamics_2009}%
  \BibitemOpen
  \bibfield  {author} {\bibinfo {author} {\bibfnamefont {J.}~\bibnamefont
  {Deschamps}}, \bibinfo {author} {\bibfnamefont {V.}~\bibnamefont {Kantsler}},
  \bibinfo {author} {\bibfnamefont {E.}~\bibnamefont {Segre}},\ and\ \bibinfo
  {author} {\bibfnamefont {V.}~\bibnamefont {Steinberg}},\ }\bibfield  {title}
  {\bibinfo {title} {Dynamics of a vesicle in general flow},\ }\href@noop {}
  {\bibfield  {journal} {\bibinfo  {journal} {Proc. Natl. Acad. Sci. U.S.A.}\
  }\textbf {\bibinfo {volume} {106}},\ \bibinfo {pages} {11444} (\bibinfo
  {year} {2009})}\BibitemShut {NoStop}%
\bibitem [{\citenamefont {Schroeder}(2018)}]{schroeder2018single}%
  \BibitemOpen
  \bibfield  {author} {\bibinfo {author} {\bibfnamefont {C.~M.}\ \bibnamefont
  {Schroeder}},\ }\bibfield  {title} {\bibinfo {title} {Single polymer dynamics
  for molecular rheology},\ }\href@noop {} {\bibfield  {journal} {\bibinfo
  {journal} {J. Rheol.}\ }\textbf {\bibinfo {volume} {62}},\ \bibinfo {pages}
  {371} (\bibinfo {year} {2018})}\BibitemShut {NoStop}%
\bibitem [{\citenamefont {Giomi}\ \emph {et~al.}(2017)\citenamefont {Giomi},
  \citenamefont {Kos}, \citenamefont {Ravnik},\ and\ \citenamefont
  {Sengupta}}]{giomi_cross-talk_2017}%
  \BibitemOpen
  \bibfield  {author} {\bibinfo {author} {\bibfnamefont {L.}~\bibnamefont
  {Giomi}}, \bibinfo {author} {\bibfnamefont {{\v{Z}}.}~\bibnamefont {Kos}},
  \bibinfo {author} {\bibfnamefont {M.}~\bibnamefont {Ravnik}},\ and\ \bibinfo
  {author} {\bibfnamefont {A.}~\bibnamefont {Sengupta}},\ }\bibfield  {title}
  {\bibinfo {title} {Cross-talk between topological defects in different fields
  revealed by nematic microfluidics},\ }\href
  {https://doi.org/10.1073/pnas.1702777114} {\bibfield  {journal} {\bibinfo
  {journal} {Proc. Natl. Acad. Sci. U.S.A.}\ }\textbf {\bibinfo {volume}
  {114}},\ \bibinfo {pages} {E5771} (\bibinfo {year} {2017})}\BibitemShut
  {NoStop}%
\bibitem [{\citenamefont {Kantsler}\ \emph {et~al.}(2007)\citenamefont
  {Kantsler}, \citenamefont {Segre},\ and\ \citenamefont
  {Steinberg}}]{kantsler_vesicle_2007}%
  \BibitemOpen
  \bibfield  {author} {\bibinfo {author} {\bibfnamefont {V.}~\bibnamefont
  {Kantsler}}, \bibinfo {author} {\bibfnamefont {E.}~\bibnamefont {Segre}},\
  and\ \bibinfo {author} {\bibfnamefont {V.}~\bibnamefont {Steinberg}},\
  }\bibfield  {title} {\bibinfo {title} {Vesicle {Dynamics} in {Time-Dependent
  Elongation Flow}: {Wrinkling Instability}},\ }\href
  {https://doi.org/10.1103/PhysRevLett.99.178102} {\bibfield  {journal}
  {\bibinfo  {journal} {Phys. Rev. Lett.}\ }\textbf {\bibinfo {volume} {99}},\
  \bibinfo {pages} {178102} (\bibinfo {year} {2007})}\BibitemShut {NoStop}%
\bibitem [{\citenamefont {Turitsyn}\ and\ \citenamefont
  {Vergeles}(2008)}]{turitsyn_wrinkling_2008}%
  \BibitemOpen
  \bibfield  {author} {\bibinfo {author} {\bibfnamefont {K.~S.}\ \bibnamefont
  {Turitsyn}}\ and\ \bibinfo {author} {\bibfnamefont {S.~S.}\ \bibnamefont
  {Vergeles}},\ }\bibfield  {title} {\bibinfo {title} {Wrinkling of {Vesicles}
  during {Transient Dynamics} in {Elongational Flow}},\ }\href
  {https://doi.org/10.1103/PhysRevLett.100.028103} {\bibfield  {journal}
  {\bibinfo  {journal} {Phys. Rev. Lett.}\ }\textbf {\bibinfo {volume} {100}},\
  \bibinfo {pages} {028103} (\bibinfo {year} {2008})}\BibitemShut {NoStop}%
\bibitem [{\citenamefont {Narsimhan}\ \emph {et~al.}(2015)\citenamefont
  {Narsimhan}, \citenamefont {Spann},\ and\ \citenamefont
  {Shaqfeh}}]{narsimhan_pearling_2015}%
  \BibitemOpen
  \bibfield  {author} {\bibinfo {author} {\bibfnamefont {V.}~\bibnamefont
  {Narsimhan}}, \bibinfo {author} {\bibfnamefont {A.~P.}\ \bibnamefont
  {Spann}},\ and\ \bibinfo {author} {\bibfnamefont {E.~S.~G.}\ \bibnamefont
  {Shaqfeh}},\ }\bibfield  {title} {\bibinfo {title} {Pearling, wrinkling, and
  buckling of vesicles in elongational flows},\ }\href
  {https://doi.org/10.1017/jfm.2015.345} {\bibfield  {journal} {\bibinfo
  {journal} {J. Fluid Mech.}\ }\textbf {\bibinfo {volume} {777}},\ \bibinfo
  {pages} {1} (\bibinfo {year} {2015})}\BibitemShut {NoStop}%
\bibitem [{\citenamefont {Schneider}\ \emph {et~al.}(2011)\citenamefont
  {Schneider}, \citenamefont {Mandre},\ and\ \citenamefont
  {Brenner}}]{schneider_algorithm_2011}%
  \BibitemOpen
  \bibfield  {author} {\bibinfo {author} {\bibfnamefont {T.~M.}\ \bibnamefont
  {Schneider}}, \bibinfo {author} {\bibfnamefont {S.}~\bibnamefont {Mandre}},\
  and\ \bibinfo {author} {\bibfnamefont {M.~P.}\ \bibnamefont {Brenner}},\
  }\bibfield  {title} {\bibinfo {title} {Algorithm for a microfluidic assembly
  line},\ }\href@noop {} {\bibfield  {journal} {\bibinfo  {journal} {Phys. Rev.
  Lett.}\ }\textbf {\bibinfo {volume} {106}},\ \bibinfo {pages} {094503}
  (\bibinfo {year} {2011})}\BibitemShut {NoStop}%
\bibitem [{\citenamefont {Leal}(2007)}]{leal_advanced_2007}%
  \BibitemOpen
  \bibfield  {author} {\bibinfo {author} {\bibfnamefont {L.~G.}\ \bibnamefont
  {Leal}},\ }\href@noop {} {\emph {\bibinfo {title} {Advanced transport
  phenomena: fluid mechanics and convective transport processes}}},\
  Vol.~\bibinfo {volume} {7}\ (\bibinfo  {publisher} {Cambridge University
  Press},\ \bibinfo {year} {2007})\BibitemShut {NoStop}%
\bibitem [{\citenamefont {Mayne}\ \emph {et~al.}(2000)\citenamefont {Mayne},
  \citenamefont {Rawlings}, \citenamefont {Rao},\ and\ \citenamefont
  {Scokaert}}]{mayne_constrained_2000}%
  \BibitemOpen
  \bibfield  {author} {\bibinfo {author} {\bibfnamefont {D.~Q.}\ \bibnamefont
  {Mayne}}, \bibinfo {author} {\bibfnamefont {J.~B.}\ \bibnamefont {Rawlings}},
  \bibinfo {author} {\bibfnamefont {C.~V.}\ \bibnamefont {Rao}},\ and\ \bibinfo
  {author} {\bibfnamefont {P.~O.}\ \bibnamefont {Scokaert}},\ }\bibfield
  {title} {\bibinfo {title} {Constrained model predictive control: {Stability}
  and optimality},\ }\href@noop {} {\bibfield  {journal} {\bibinfo  {journal}
  {Automatica}\ }\textbf {\bibinfo {volume} {36}},\ \bibinfo {pages} {789}
  (\bibinfo {year} {2000})}\BibitemShut {NoStop}%
\bibitem [{\citenamefont {Houska}\ \emph {et~al.}(2011)\citenamefont {Houska},
  \citenamefont {Ferreau},\ and\ \citenamefont {Diehl}}]{houska_acado_2011}%
  \BibitemOpen
  \bibfield  {author} {\bibinfo {author} {\bibfnamefont {B.}~\bibnamefont
  {Houska}}, \bibinfo {author} {\bibfnamefont {H.~J.}\ \bibnamefont
  {Ferreau}},\ and\ \bibinfo {author} {\bibfnamefont {M.}~\bibnamefont
  {Diehl}},\ }\bibfield  {title} {\bibinfo {title} {{ACADO} toolkit---{An}
  open-source framework for automatic control and dynamic optimization},\
  }\href@noop {} {\bibfield  {journal} {\bibinfo  {journal} {Optim. Contr.
  Appl. Met.}\ }\textbf {\bibinfo {volume} {32}},\ \bibinfo {pages} {298}
  (\bibinfo {year} {2011})}\BibitemShut {NoStop}%
\bibitem [{\citenamefont {Quirynen}\ \emph {et~al.}(2015)\citenamefont
  {Quirynen}, \citenamefont {Vukov}, \citenamefont {Zanon},\ and\ \citenamefont
  {Diehl}}]{quirynen_autogenerating_2015}%
  \BibitemOpen
  \bibfield  {author} {\bibinfo {author} {\bibfnamefont {R.}~\bibnamefont
  {Quirynen}}, \bibinfo {author} {\bibfnamefont {M.}~\bibnamefont {Vukov}},
  \bibinfo {author} {\bibfnamefont {M.}~\bibnamefont {Zanon}},\ and\ \bibinfo
  {author} {\bibfnamefont {M.}~\bibnamefont {Diehl}},\ }\bibfield  {title}
  {\bibinfo {title} {Autogenerating microsecond solvers for nonlinear {MPC}: a
  tutorial using {ACADO} integrators},\ }\href@noop {} {\bibfield  {journal}
  {\bibinfo  {journal} {Optim. Contr. Appl. Met.}\ }\textbf {\bibinfo {volume}
  {36}},\ \bibinfo {pages} {685} (\bibinfo {year} {2015})}\BibitemShut
  {NoStop}%
\bibitem [{\citenamefont {Qin}\ \emph {et~al.}(2010)\citenamefont {Qin},
  \citenamefont {Xia},\ and\ \citenamefont {Whitesides}}]{qin_soft_2010}%
  \BibitemOpen
  \bibfield  {author} {\bibinfo {author} {\bibfnamefont {D.}~\bibnamefont
  {Qin}}, \bibinfo {author} {\bibfnamefont {Y.}~\bibnamefont {Xia}},\ and\
  \bibinfo {author} {\bibfnamefont {G.~M.}\ \bibnamefont {Whitesides}},\
  }\bibfield  {title} {\bibinfo {title} {Soft lithography for micro-and
  nanoscale patterning},\ }\href@noop {} {\bibfield  {journal} {\bibinfo
  {journal} {Nat. Protoc.}\ }\textbf {\bibinfo {volume} {5}},\ \bibinfo {pages}
  {491} (\bibinfo {year} {2010})}\BibitemShut {NoStop}%
\end{thebibliography}


\end{document}